%% file: barriermc.tex
\documentclass[11pt,a4paper]{amsart}

\usepackage{amsmath,amssymb,latexsym,amsthm,enumerate}
\usepackage{verbatim}
\usepackage{graphicx} 
\usepackage{hyperref} 
\usepackage{xcolor} 
\usepackage{psfrag} 

\def\E{\mathbb{E}}
\def\EE{\mathbf{E}}

\def\NN{\mathbb{N}}

\def\RR{\mathbb{R}}

\def\PP{\mathbb{P}}

\def\GG{\mathbb{G}}
\def\P{\mathbf{P}}
\def\L{\mathcal{L}}
\def\A{\mathcal{A}}

\def\G{{\mathcal G}}

\def\1{\mathbf{1}}
\def\0{\mathbf{0}}

\def\I{\mathbf{I}}

\newcommand{\tL}{\widehat{\L}}
\newcommand{\te}{\widehat{e}}

\newtheorem{Thm}{Theorem}

\newtheorem{Lemma}{Lemma}

\newtheorem{Def}{Definition}
\newtheorem{Ex}{Example}
\newtheorem{As}{Assumption}
\newtheorem{Rem}{Remark}

\def\BEN{\begin{enumerate}}  \def\BI{\begin{itemize}}
\def\EEN{\end{enumerate}}   \def\EI{\end{itemize}}
    \def\nn{\nonumber}

\def\mbb{\mathbb} \def\mbf{\mathbf} 
\def\mc{\mathcal} \def\unl{\underline} \def\ovl{\overline}

\def\le{\left}
\def\ri{\right}
\def\te#1{\mathrm{e}^{#1}}   
\def\WT{\widetilde}
\def\WH{\widehat}

    \def\a{\alpha}

\def\e{\epsilon}
 
  \def\nn{\nonumber}   
     
 \def\q{\qquad} 
 \def\L{\Lambda}  
  \def\td{\text{\rm d}}
 
\numberwithin{equation}{section}

\newcommand{\exit}{{\mbox{\, \vspace{3mm}}}
\hfill\mbox{$\square$}}

\newenvironment{pr}{\vspace{1mm}\noindent\textbf{Proof.}}
                   {\vspace{-5mm}\begin{flushright}$\Box$\end{flushright}}

\begin{document}

\title{Continuously monitored barrier options under Markov processes}

\author{Aleksandar Mijatovi\'{c}}
\address{Department of Statistics, University of Warwick}
\email{a.mijatovic@warwick.ac.uk}
\author{Martijn Pistorius}

\thanks{{\it Acknowledgements:} 
We would like to thank the anonymous referees, 
Bjorn Eriksson, Mike Giles, Vassili Kolokoltsov, Steven Kou, Sergei Levendorskii, 
Dilip Madan, Jan Obloj and Johannes Stolte 
as well as the participants of the 
2009 Leicester workshop on Spectral and Cubature Methods 
in Finance and Econometrics 
for useful suggestions and constructive comments, 
which led to improvements of the paper.
Research supported by EPSRC grant {EP/D039053}.
This research was carried out while AM was based at Imperial College London.
}

\address{Department of Mathematics, Imperial College London}
\email{m.pistorius@imperial.ac.uk}

\begin{abstract}
In this paper we present an algorithm for pricing barrier options in
one-dimensional Markov models. The approach rests on the construction
of an approximating continuous-time
Markov chain that closely follows the dynamics
of the given Markov model. We illustrate the method by implementing
it for a range of models, including a local 
L\'evy process and a local volatility jump-diffusion.
We also provide a convergence proof and error estimates for this algorithm.
\end{abstract}

\maketitle

\section{Introduction}
\label{sec:Intro}
\input{intro.tex}

\section{Problem setting: Barrier options for Markov processes}
\label{sec:barrier_MP}

\input{barrier_mp2.tex}

\section{Exit probabilities for continuous-time Markov chains}
\label{sec:Barriers_MC}
\input{barriersforchains1.tex}

\section{Construction of the Markov chain}
\label{sec:MarkovChainConstr}
\input{construction.tex}

\section{Convergence and error estimates}
\label{sec:WeakConv}
\input{convergence.tex}

\section{Numerical examples}
\label{sec:numericalEx}
\input{numericalex1.tex}

\section{Conclusion}
\label{sec:conclusion}
\input{conclusion.tex}

\appendix
\section{Proofs}
\label{sec:proofs}
\input{proofs.tex}

\bibliographystyle{plain}
\bibliography{cite}

\end{document}

%% file: intro.tex
\subsection{Background and motivation}

Barrier options are among the most popular exotic derivatives. 
Such contracts form effective risk management tools, and are 
liquidly traded in the Foreign Exchange markets.
The most liquid barrier options in FX markets
are continuously monitored
single- or double-no-touch options and knock-in or 
knock-out calls and puts 
(see e.g. Hakala and Wystup~\cite{FX_Risk},
Lipton~\cite{Lipton_Book}, \cite{Lipton_Universal}, 
Wystup~\cite{Wystup_FX}).
The main challenge 
in the risk management of large portfolios of barrier 
options 
faced by trading desks that make a market in these securities
is to be able to price and hedge the barrier products in models
that are flexible enough to describe the observed option prices
(i.e. calibrate to the vanilla market price quotes).

It is by now well established that the classical Black-Scholes 
model lacks the flexibility to fit accurately to observed option
price data
(see e.g. Gatheral~\cite{Gatheral} and the references therein).
A variety of models have been proposed to provide an improved 
description of the dynamics of the price of the underlying 
that can more accurately describe the option surface.
Parametric diffusion models like the CEV process~\cite{Cox_CEV} 
have additional flexibility to fit the vanilla skew at a single
maturity for as many options as there are free parameters in the 
model. The seminal idea (developed by Dupire~\cite{Dupire_Smile} and 
Gy\"ongy~\cite{Gyongy}) 
that allows one to construct a model that
can describe the entire implied volatility surface (across all strikes and
maturities) is that of local volatility models, where a non-parametric
form of the local volatility function is constructed from the option
price data. It has been shown that in practice such models imply unrealistic
dynamics of the option prices (see the formula for the implied volatility in a local 
volatility model given in~\cite{Hagan_LocalVol}). The ramification 
is an unrealistic amount of vega risk, which is expensive to hedge.
Therefore, even though in a local volatility model barrier
options can be priced using a PDE solver, this 
modelling framework alone 
is not suitable for the risk management of a large portfolio of barrier 
options.

At the other end of the spectrum are the jump processes 
with stationary and 
independent increments, which can fit very well the volatility smile at 
a single maturity (see e.g.~\cite{ContTankov} and the references 
therein).  A variety of models in the exponential L\'{e}vy 
class have been proposed in the literature: CGMY~\cite{CarrMadanGemanYor_CGMY}, 
KoBoL~\cite{TheBook}, 
generalised hyperbolic~\cite{Eberlein_Hyperbolic}, NIG~\cite{NIG} and Kou~\cite{Kou}. 
Exponential L\'evy processes are simple examples of Markov processes whose law is uniquely
determined by the distribution of the process at a single time. 
Since the set of call option prices at a 
fixed maturity  for all strikes uniquely determines the marginal 
risk-neutral distribution at that maturity, 
calibration to option prices at multiple maturities in principle 
fixes the corresponding marginals. It has been reported (see e.g. \cite{ContTankov}) that L\'{e}vy processes lack the flexibility of calibrating simultaneously across a range of strikes and maturities. 
Several generalisations within 
the one-dimensional Markov framework 
have been proposed. 

If the stationarity assumption is relaxed 
while the property of independence of increments is retained,
one arrives at the class of exponential additive processes,
which have recently been shown to calibrate well to several maturities 
in equity markets. 
The Sato process introduced in Carr et al.~\cite{CarrMadanGemanYor_selfdec}
is an example of such an additive model used in financial modelling.

The independent increments property of a process implies that its 
transition probabilities are translation invariant in space, so that 
they only depend on the difference between the end and starting value
of the process. 
It is well known that the distribution 
of a log-asset price depends in a non-linear
way on the starting point (e.g. in equity markets it has been observed
that if the current price is high, then the volatility is low and vice versa).
To capture this effect one is led to consider Markov 
jump-processes whose increments are not independent.
As a generalisation of local volatility models,
the class of local L\'evy processes introduced by 
Carr et al.~\cite{CarrMadanGemanYor_LocalLevy} 
allows the modeller to modulate the intensity of the jumps as well as their
distribution depending on where the underlying asset is trading. 
A local volatility jump-diffusion with similar structural 
properties was 
calibrated to the implied volatility surface
in Andersen and Andreasen~\cite{Andersen_Andreasen_Jump_Diff}
and He et al.~\cite{Jump_Diff_Callibration}.
%
Due to the presence of jumps and the absence of 
stationarity and independence of increments,
the problem of obtaining the first-passage probabilities 
for such a general class of processes is 
computationally less tractable. 

There exists currently a good deal of literature on numerical methods 
for the pricing of barrier-type options. 
It is well known that in this case a straightforward 
Monte Carlo simulation algorithm
will be time-consuming and yield unstable results for the prices 
and especially the sensitivities. The knock-in/out features
in the barrier option payoffs lead to slower convergence
of the Monte Carlo algorithm.
To address this problem the following (semi-)analytical approaches 
have been developed for specific models:
\begin{enumerate}
\item[(a)] spectral expansions for several parametric diffusion models 
(Davydov and Linetsky \cite{DavydovLinetsky_Eig}, Lipton~\cite{Lipton_Book}),
\item[(b)] transform based approaches for exponential L\'{e}vy models
(Boyarchenko and Levendorskii~\cite{BoyLev_doublebarrier}, 
Geman and Yor~\cite{GemanYor}, 
Jeannin and Pistorius~\cite{Jeannin:2008},
Kou and Wang~\cite{Kou_Wang}, Sepp~\cite{Sepp}),
\end{enumerate}
The method 
(a) employs the explicit spectral decompositions 
for this class of diffusion models,
whereas the approach
(b) exploits the independence and stationarity 
of the increments of the L\'{e}vy process,
and the so-called Wiener-Hopf factorisation. 
Since both of these approaches hinge on special structural 
properties of the underlying processes,
it is not clear if and how they can be extended to
 more general Markovian models.

A different approach, pioneered by Kushner (see e.g. \cite{Kushner2}), 
is the discrete time Markov chain approximation method. 
Originally developed for the numerical 
solution of stochastic optimal control problems in continuous time, 
this method consists of approximating the system of interest 
by a discrete time chain that closely follows its dynamics, and solving the problem of interest for this chain.  
An application to the pricing of American type options 
is given in Kushner \cite{Kushner1}. Bally et al. \cite{GillesP} 
develop a quantization method to efficiently value American options 
on baskets of assets under a local volatility model.
Using a discrete time Markov chain, Duan et al. \cite{Duanetal} price 
a discretely monitored barrier option in the Black-Scholes 
and NGARCH models. Rogers and Stapleton \cite{RS} investigate an 
efficient binomial tree method for 
barrier option pricing (see also references therein for related methods).
Related are PDE and PIDE finite difference 
discretization methods that have been investigated by various authors;
Zvan et al. \cite{Zvan97pdemethods} consider 
barrier and related options in the Black-Scholes model; 
Tavella and Randall~\cite{TavellaRandall} present an overview of PDE finite difference methods
for the pricing of financial instruments;
Wang et al. \cite{Forsyth_CGMY} develop a robust scheme for American options under a CGMY model, and Cont and Voltchkova \cite{ContV} 
follow a viscosity approach for the PIDEs connected to European 
and barrier option under L\'{e}vy models. Markov chains have also been employed to directly model the evolution of price processes; 
Albanese and Mijatovi\'c \cite{AM} model the stochasticity of  
risk reversals and carried out a 
calibration study in FX markets under a certain 
continuous-time Markov chain constructed to model the 
FX spot process.

%


\subsection{Contribution of the current paper}

\begin{figure}[t]
\centering
\psfrag{G1}{$\widehat{\mathbb{G}}$} 
\psfrag{G2}{$\widehat{\mathbb{G}}$}
\psfrag{L11}{\Large $\Lambda_{11}$} 
\psfrag{L12}{\Large $\Lambda_{12}$} 
\psfrag{L21}{\Large $\Lambda_{21}$} 
\psfrag{L22}{\Large $\Lambda_{22}$} 
\psfrag{L}{\huge $\Lambda$} 
\psfrag{Lt}{\huge $\widetilde{\Lambda}$} 
\psfrag{011}{\Large $0$} 
\psfrag{012}{\Large $0$} 
\psfrag{Lh}{\huge $\widehat{\Lambda}$} 
\psfrag{L22h}{\Large $\Lambda_{22}$} 
\includegraphics[width=0.7\textwidth]{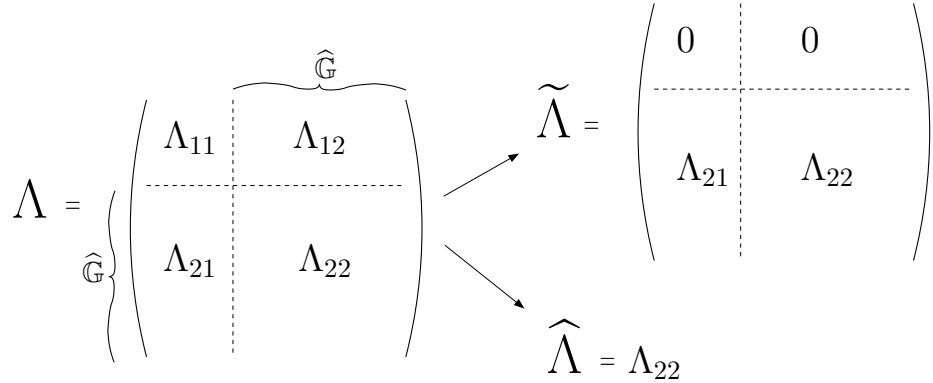}
\caption{\footnotesize This is a schematic picture 
in bloc notation
that 
demonstrates how the matrices 
$\WH \Lambda$
and
$\widetilde{\Lambda}=\widetilde{\Lambda}_0$
are obtained from the generator
$\Lambda$.
The matrix 
$\Lambda$
is the generator matrix of an approximating continuous-time 
Markov chain
$X$
on the state-space
$\GG$.
The subset
$\WH\GG\subset \GG$
consists of the elements of 
$\GG$
that lie between the barriers.
The matrix
$\WH \Lambda$
contains all the information necessary to price any contract 
that knocks out (or knocks in) when a barrier is breached.
Similarly the 
matrix
$\widetilde{\Lambda}$
is what is required to compute the distribution of any function 
that depends on the first-passage and
overshoot of the chain
$X$
into the region on the other side of the barriers.
}
\label{fig:MartixOp}
\end{figure}

In this paper we consider the problem of pricing 
barrier options in the setting of one-dimensional Markov processes, 
which in particular includes the case of local L\'evy models as well as
local volatility jump-diffusions 
and additive processes. The presented approach is probabilistic in nature and 
is based on the following two elementary observations:
(i) given a Markov asset price process 
$S$ it is straightforward to construct 
a continuous-time Markov chain model $X$
whose law is close to that of $S$,   
by approximating the generator of the process $S$
with an intensity matrix; (ii)
the corresponding first-passage problem for a continuous-time Markov chain
can be solved explicitly via a closed-form formula 
that only involves the generator matrix of the chain
$X$. 
More precisely, for a given Markov asset price model 
$S$ 
on the state-space $\E=[0,\infty)$ with corresponding 
generator $\mathcal{L}$
the algorithm for the pricing of any barrier product (including 
rebate options, which depend on the position at the moment of first-passage)
consists of the following two steps:

\begin{enumerate}
\item[(i)] Construct a finite state-space
$\GG\subset\E$
and a generator matrix
$\L$
for the chain
$X$
that approximates the operator
$\mathcal{L}$
on $\GG$.
\item[(ii)] To value knock-out and 
rebate options, obtain the matrices 
$\WH \L$ and $\widetilde{\Lambda}$
by the procedure in Figure~\ref{fig:MartixOp}
and apply closed-form formulas in 
terms of 
these matrices
(given in equations~\eqref{eq:BarrierPrice} and \eqref{eq:BarrierPrice2}
below).
\end{enumerate}
The form of the generators of  Markov processes that commonly 
arise in pricing theory (including the local L\'evy class) 
is well known from general theory (and are reviewed 
in Section~\ref{sec:barrier_MP}). 
The state-space $\mathbb G$ in step (i) is taken to be non-uniform with a higher density of points in the relevant areas
such as spot and barrier levels. 
Subsequently, the generator matrix $\Lambda$ is defined by 
matching the instantaneous moments of the 
Markov processes
$S$
and the chain
$X$
on the state-space
$\GG$.
This criterion implies in particular that the chain 
$X$ has the same average drift as the asset price process
$S$.
Step (ii) of the algorithm consists of the evaluation
of the closed-form formulas for the first-passage probabilities that 
can be derived employing continuous-time Markov chain theory (see Theorem~\ref{thm:Main}
below).
The evaluation of this formula consists of exponentiation
of either the matrix
$\WH \L$
or 
$\widetilde{\L}$,
which can be performed using the Pad\'e approximation algorithm
that is implemented in standard packages such as Matlab (see also~\cite{Higham}).
The outputs of the algorithm yield arbitrage-free prices 
in a certain continuous time Markov chain 
model for the risky asset price process. 
This feature is a consequence
of the fact that the algorithm is based on the construction 
of an approximate stochastic model.
This property is not 
shared by many other numerical methods used in practice 
that are based on 
purely analytical considerations.

We implemented this algorithm for a number of models
that include the features of local volatility as well as 
jumps. 
We obtained an accurate match with the numerical results 
under diffusion and L\'{e}vy models that were considered 
elsewhere in the literature (see Section~\ref{sec:numericalEx}). 
We prove that by refining the grid the prices 
generated by this approach converge to those of the limiting model. 
We also establish, under additional 
regularity assumptions, an error bound that is linear in the spatial mesh size and the truncation error 
(see Section~\ref{sec:WeakConv}). 
We showed that an additional logarithmic factor may arise in this error bound
when the L\'evy density  
has a pole of order two at the origin.
Numerical experiments (reported in Section~\ref{sec:numericalEx})
appear to suggest that, for a number of models, the error actually decays
quadratically in the spatial mesh size. 
An extension to the case of time-dependent characteristics is presented 
in an unabridged version of this paper \cite{MP_2010_LongVersion}, which also includes extra numerical examples, as well as some sample code.

There is good deal of literature devoted to the study of the (weak) 
convergence of Markov chains to limiting processes. 
However the (sharp) rates of convergence of prices generated by the Markov chain approximation
to those of the limiting model are rarely available, 
especially for barrier options.
For discrete time Markov chains some explicit rates have been established
(see e.g. Broadie et al.~\cite{BGK} and Gobet and Menozzi~\cite{Gobet}).
Establishing the sharp rates of convergence 
for specific models remains an 
open question, left for future research. 

The remainder of the paper is organized as follows. 
In Section~\ref{sec:barrier_MP}
we define  the class of models and barrier option 
contracts that is considered, and
state some preliminary results about Markov processes.
Section~\ref{sec:Barriers_MC} presents the formulas for the
first-passage quantities of the continuous-time Markov chains.
In Section~\ref{sec:MarkovChainConstr}
we describe the discretization algorithm to construct
the intensity matrix
$\Lambda$
of the chain 
$X$. Section~\ref{sec:WeakConv}
states the convergence results, which are proved in
Appendix~\ref{sec:proofs}.
Numerical results are presented in
Section~\ref{sec:numericalEx}
and Section~\ref{sec:conclusion}
concludes the paper.

%% file: barrier_mp2.tex
The problem under consideration is that of
the valuation of general barrier options, which can be formulated as
follows. Given a random process $S=\{S_t\}_{t\ge 0}$ modelling the
price evolution of a risky asset, non-negative payoff and rebate functions
$g$ and $h$, and a set $A$ specifying the range of values for which the
contract `knocks out', it is of interest to evaluate the expected
discounted value of the random cash flow associated to a general 
barrier option contract
\begin{equation}\label{eq:barrierpayoff}
g(S_T) \I_{\{\tau_A > T\}} + h(S_{\tau_A})\I_{\{\tau_A \leq T\}},
\end{equation}
where $\I_C$ denotes the indicator of a set $C$ and
$$\tau_A = \inf\{t\ge 0: S_t\in A\}$$ 
is the first time that $S$ enters the set $A$. Furthermore, 
it is relevant to quantify the  
sensitivities of this value with respect to different parameters 
such as the spot value $S_0=x$. 
The cash flow in~\eqref{eq:barrierpayoff} 
consists of a payment $g(S_T)$ in the case 
the contract has not knocked out by the time 
$T$, 
and a 
rebate $h(S_{\tau_A})$ if it has. 
Examples of commonly traded options included in this setting  
are the down-and-out, up-and-out and double knock-out options.
In particular, by taking $A=\emptyset$ we
retrieve the case of a standard European claim with payoff
$g(S_T)$ at maturity $T$.

We will consider this valuation problem in a Markovian setting,
assuming that the underlying $S$ is a Markov process with
state-space $\E:=[0,\infty)$ 
defined on some filtered probability
space $(\Omega,\mathcal F, \mbf F , \P)$, where $\mbf F = \{\mc F_t\}_{t\ge 0}$ denotes the standard filtration generated by
$S$. Thus, $S$ takes values in
$\E$ and satisfies the Markov property:
\begin{equation}\label{eq:markovdef}
\EE[f(S_{t+s})|\mc F_t] = P_sf(S_t),
\end{equation}
for all $s,t\ge 0$ and bounded 
Borel functions $f$, where $\EE$ denotes the expectation under the 
probability measure $\P$
and $P_sf$ is given by 
\begin{equation}\label{eq:Pf}
P_sf(x) := \EE_x[f(S_s)] := \EE[f(S_s)|S_0=x]. 
\end{equation}
By taking
expectations in \eqref{eq:markovdef} we see that the family $(P_t)_{t\ge 0}$
forms a semigroup: 
\begin{equation*}
P_{t+s}f = P_t(P_sf),\quad \text{for all $s,t \ge0$, and $P_0f=f$.}
\end{equation*}
Informally, these conditions state that the expected value of the random cash flow $f(S_{t+s})$ occurring at time 
$t+s$ 
conditional on the available information up to time 
$t$ 
depends on the past via the value 
$S_t$ 
only. Setting the rate 
of discounting equal to a non-negative constant 
$r$, 
for any pair of non-negative Borel functions 
$g$ and $h$
the expected discounted value of the barrier cash flow
\eqref{eq:barrierpayoff} 
at the epoch $\tau_A\wedge T$,
the earlier of maturity $T$ and the first entrance time $\tau_A$,
is given by
\begin{equation}\label{eq:PTA}
\EE_x\le[\te{-rT} g(S_T) \I_{\{\tau_A > T\}}\ri] +
\EE_x\le[\te{-r\tau_A}h(S_{\tau_A})\I_{\{\tau_A \leq T\}}\ri].
\end{equation}
If $S$ represents the price of a tradeable asset, 
$r$ 
is the
risk-free rate,
$d$
is the dividend yield and the process
$\{\te{-(r-d)t}S_t\}_{t\ge 0}$ is a martingale, 
standard arbitrage
arguments imply that 
no arbitrage is introduced if 
expression~\eqref{eq:PTA}
is used as the current price of the option 
with payoff \eqref{eq:barrierpayoff}. 

Before proceeding we review some key concepts of the 
standard Markovian setup that will be needed in the sequel. 
For background on the (general) theory of Markov processes 
we refer to 
the classical works Chung and Walsh \cite{ChungWalsh}, 
Ethier and Kurtz~\cite{EthierKurtz}, 
It\^o and McKean \cite{ItoMcKean} and Rogers and 
Williams \cite{RogersWilliams} 
(the latter two in particular treat the case of 
Markov processes with continuous sample paths). 
In what follows we will 
restrict 
$S$ 
to be in a subclass of Markov 
processes for which,
if the function
$f$
is continuous and tends to 
zero at infinity,
the expected payoff 
$P_tf(x)$ 
has the following  properties:
it
depends continuously on
the spot
$S_0=x$
and on expiry
$t$
and also decays to zero when 
$x$
tends 
to infinity.
More precisely, denoting by $C_0(\E)$ the set of continuous functions 
$f$ on $\mathbb E$ that tend to zero at infinity,
we make the following assumption:
\begin{As}\label{def:Feller}
$S$ is a {\em Feller process} on $\E$, that is, 
for any $f\in C_0(\E)$, the family $(P_tf)_{t\ge 0}$,
with $P_tf$ defined in \eqref{eq:Pf},
satisfies:
\begin{itemize}
\item[(i)] $P_tf \in C_0(\E)$ for any $t>0$;
\item[(ii)] $\lim_{t\to 0}P_tf(x) = f(x)$ for any $x\in \E$.%
\end{itemize}
\end{As}
The Feller property guarantees 
that there exists a version of the process $S$ 
with c\`adl\`ag paths satisfying the strong Markov property. 
In particular, a Feller process is a Hunt process. 

Throughout the paper we will take the knock-out set $A$ to be 
of the form
\begin{equation}\label{eq:A}
A = [0,\ell]\cup[u,\infty),
\qquad 0\leq\ell<u\leq\infty,
\end{equation}
which includes the cases of double and 
single barrier options---the latter 
by taking $\ell=0$ or $u=\infty.$
To rule out degeneracies we will  
make the following assumption on the behaviour of $S$ 
at the boundary points $\ell$ and $u$:

\begin{As}\label{as1}
For every 
$x\in\mbb{E}$
we have
$\P_x(\tau_A = \tau_{A^o}) = 1$ where $A^o = [0,\ell)\cup(u,\infty)$.
\end{As}

This assumption states that the first entrance times into 
$A$
and its interior coincide almost surely, if the spot $S_0$ is equal to $x$.
If 
$u<\infty$,
a sufficient condition for Assumption \ref{as1} to be
satisfied is $\P_x(\tau_{A^o}=0)=1$ for $x\in\{\ell, u\}$; 
that is, when started at
$\ell$
or
$u$,
the process 
$S$
immediately enters 
$A^o$.

The class of Feller processes satisfying Assumption~\ref{as1} 
includes many of the 
models employed 
in quantitative finance such as (Feller-)diffusions, jump-diffusions with non-generate diffusion coefficient
and L\'{e}vy processes whose L\'{e}vy measure admits a density.

The family $(P_t)_{t\ge 0}$ is determined by its infinitesimal
generator $\mathcal L$ that is defined as
\begin{equation}\label{eq:Lf}
\mc Lf(x) := \lim_{t\downarrow 0} \mbox{$\frac{1}{t}$}(P_tf - f)(x)
\end{equation}
for any function 
$f\in C_0(\E)$
for which the right-hand side of~\eqref{eq:Lf}
converges in the strong sense.\footnote{That is, the 
convergence is with respect to the norm $\|f\|:=\sup_{x\in \E}|f(x)|$ 
of the Banach space $(C_0(\E),\|\cdot\|)$.} 
The set 
$\mc D$ 
of such functions 
is called the domain of the operator 
$\mathcal{L}$
and is dense in 
$C_0(\E)$.
These fundamental facts 
about semigroups and their generators can be found in~\cite[Ch. 1]{EthierKurtz}.

We next give a few examples of Feller processes with their generators. 
\begin{Ex}\label{ex:dif}\rm 
A diffusion asset price model $S=\{S_t\}_{t\geq0}$
evolves under a risk-neutral measure according to
the stochastic differential equation (SDE)
\begin{eqnarray}
\label{eq:dif}
\frac{\td S_t}{S_t} =  \gamma \td t+\sigma\left(S_t\right)\td W_t,
\end{eqnarray}
where $S_0>0$
is the initial price, $\gamma\in\RR$ and $\sigma:\mathbb R_+\to\RR_+$ 
is a given measurable function. To guarantee the absence of arbitrage 
we assume that $\sigma$ is chosen such that 
the discounted process $\{\te{-\gamma t}S_t\}_{t\geq0}$ is a martingale,
If, in addition, infinity is not entrance\footnote{
See It\^o and McKean~\cite{ItoMcKean} for an explicit criterion in terms of $\gamma$ and $\sigma$ 
for this to be the case.} for $S$
and $\sigma$ is a continuous function, then
$S$ is a Feller process, 
and its infinitesimal generator $\mathcal L_D$ acts on 
$f\in C^2_c(\mathbb E)$\footnote{$C^2_c(\mathbb E)$ denotes the 
set of $C^2$ functions with compact support in 
$\mathbb E^o=(0,\infty)$.} as 
\begin{equation}\label{LD}
\mathcal L_Df(x) = \frac{\sigma(x)^2x^2}{2}f''(x) + \gamma x f'(x),
\end{equation}
where $f'$ 
denotes the derivative 
of $f$ with respect to $x$ (see~\cite[Sec. 8.1]{EthierKurtz}).
\end{Ex}

\begin{Ex}\rm
The price process in an exponential L\'{e}vy model
$S$ 
given by
\begin{equation}
\label{eq:Levy}
S_t := S_0\te{(r-d) t}  \frac{\te{L_t}}{\EE[\te{L_t}]}
\end{equation}
where $r$ and $d$ are constants representing 
the interest rate and dividend yield and 
$L=\{L_t\}_{t\ge 0}$ is a L\'{e}vy process, 
such that $\EE[\te{L_t}]<\infty$ 
for all $t>0$. By construction, $\{\te{-(r-d)t}S_t\}_{t\geq0}$ is a martingale. Further, $\EE[\te{L_t}]<\infty$ if and only if the L\'{e}vy measure 
$\nu$ integrates $\exp(y)$ at infinity, that is,
\begin{eqnarray}
\label{eq:Levy_Measure_Cond}
\int_{(1,\infty)}\te{y} \nu(\td y)<\infty.
\end{eqnarray}
The law of $L$ is determined by its characteristic 
exponent 
$\Psi$, 
which is related to the characteristic 
function 
$\Phi_t$ of $L_t$ by $\Phi_t(s) = \exp(t\Psi(s))$  
and which, under condition \eqref{eq:Levy_Measure_Cond}, 
has the L\'{e}vy-Khintchine representation
$$
\Psi(s) = {\mathtt i} cs - \frac{\sigma^2s^2}{2} + \int_{\mbb R} \le(\te{\mathtt{ i} sy} - 1 - \mathtt{i} s y\ri)\nu(\td y),
$$
where $(c,\sigma^2,\nu)$ is the characteristic triplet, with $\sigma, c\in\mathbb R$ and $\nu$ the L\'{e}vy measure, which satisfies
the integrability condition $\int_{(-1,1)}|y|^2\nu(\td y) <\infty$.
The process $S$ is a Feller process with an infinitesimal generator acting 
on $f\in C^2_c(\mathbb E)$ as (cf. Sato \cite[Thm. 31.5]{Sato})
\begin{equation*}
\mc Lf(x) = \frac{\sigma^2x^2}{2}f''(x) + \gamma x f'(x) + \int_{\mathbb R}[f(x\te{y}) 
- f(x) - x f'(x)(\te{y}-1)]\nu(\td y),
\end{equation*}
where 
$\gamma = r -d.$
\end{Ex}

\begin{Ex}\rm More generally, one may specify the 
price process $S$ by 
directly prescribing its 
generator $\mathcal L$ 
to act on sufficiently 
regular functions $f$ as
\begin{equation}\label{eq:gengen}
\mathcal L f(x) = \mc L_Df(x) + \mc L_Jf(x),
\end{equation}
where $\mc L_Df$ is given in \eqref{LD} and
\begin{equation}
\mc L_Jf(x) = \int_{(-1,\infty)}[f(x(1+y)) - f(x) - f'(x)xy]\nu(x,\td y),\nn
\end{equation}
where for every $x\in \E$, 
$\nu(x,\td y)$ is a (L\'{e}vy) measure with 
support in $(-1,\infty)$ such 
that 
$$
\int_{(-1,\infty)}\min\{ y^2, |y| \} \nu(x,\td y) < \infty.
$$ 
The discounted process $\{\te{-\gamma t}S_t\}_{t\geq0}$ 
is a local martingale. Sufficient conditions on $\sigma$ and $\nu$ 
to guarantee the existence of a Feller process $S$ corresponding to
this generator were established 
in  Kolokoltsov~\cite[Thm. 1.1]{Kolokoltsov}. 
\end{Ex}

A key step in solving the barrier valuation problems is the observation that the expected values of knock-out options and general barrier options 
can be expressed in terms of the {\em marginal} distributions of 
two Markov processes associated to $S$. Given a Markov process $S$ 
the processes that have the same dynamics as $S$ before entering $A$, but
are stopped or jump to the graveyard state $\partial$ upon entering 
the set $A$, respectively, are themselves Markov processes. 
Let
$\WH S^A=\{S_t\,\I_{\{t<\tau_A\}} 
+ \partial\,\I_{\{t\ge\tau_A\}}\}_{t\ge 0}$ 
denote the killed process and  
$\WT S^A$
the process killed at rate
$r$
that is stopped upon entering the set 
$A$.
Then the value of the barrier options can be expressed as
\begin{eqnarray*}
\EE_x\le[g(S_T) \I_{\{\tau_A > T\}}\ri] 
= \EE_x\le[g(\WH S_T^A)\ri] 
&=:&  \WH P^A_Tg(x),\\
\EE_x\le[\te{-rT} g(S_T) \I_{\{\tau_A > T\}}\ri] +
\EE_x\le[\te{-r\tau_A}h(S_{\tau_A})\I_{\{\tau_A \leq T\}}\ri]
&=\phantom{:}& 
\EE_x[\te{-r(T\wedge \tau_A)}f(S_{T\wedge\tau_A})]\\
&=\phantom{:}&\EE_x[f(\WT S_{T}^A)] =: \WT P^A_Tf(x),
\end{eqnarray*}
where we assume that $g(\partial)=0$ 
and the function $f$ is defined as 
$f(x)=\I_A(x) h(x)+\I_{\RR_+\backslash A}(x) g(x)$.
To calculate the value-functions of barrier options written 
on the underlying price process $S$ we thus need to identify 
$\WH P^A_Tg(x)$ and 
$\WT P^A_Tf(x)$. This can be achieved by employing the 
infinitesimal generators of the semigroups 
$(\WT P^A_t)_{t\ge 0}$ and $(\WT P^A_t)_{t\ge 0}$ associated
to the Markov processes $\WT S^A$ 
and $\WH S^A$ which are explicitly expressed
in terms of the generator $\mathcal L$ as follows:

\begin{Lemma}\label{lem:KeyLemma} 
(i) For any $f\in\mc D$, where $\mc D$
is the domain of the generator $\mathcal{L}$, 
we have
\begin{eqnarray}\label{eq:LA}
\lim_{t\downarrow 0} t^{-1}(\WT P_t^Af(x) - f(x)) &=& \WT k(x)
:= \begin{cases}
0,& x\in A,\\
(\mc{L}-r)f(x), & x\notin A,
\end{cases}
\end{eqnarray}
where the convergence is pointwise. If $\WT S^A$ is a
Feller process, and 
\begin{equation}\label{eq:boundary}
\lim_{x\to\partial A}\le\{\mc Lf(x) - rf(x)\ri\}=0, \q\text{where $\partial A$ is the boundary of $A$},
\end{equation}
then $\WT{\mc L}^Af=\WT k$,  
where $\WT{\mc L}^A$ is the 
infinitesimal generator of the semigroup 
$(\WT P^A_t)_{t\geq0}$.

(ii) Let $g\in\mathcal D$ and assume that $x$ satisfies
\begin{equation}\label{eq:hatcond}
\P_x(\tau_A\leq t)=o(t)\>\text{ as }\>t\searrow0
\q\text{or}\q g(S_{\tau_A})=0\q \P_x\text{-a.s.}
\end{equation}
Then
\begin{eqnarray}
\label{eq:LA2}
\lim_{t\downarrow 0} t^{-1}(\WH P_t^Ag(x) - g(x)) &=& \mc{L}g(x), 
\end{eqnarray}
where the convergence is pointwise. If $\WH S^A$ is a
Feller process and 
$$
g|_A = 0\q\text{and}\q\lim_{x\to\partial A}\mc Lg(x)=0,
$$
then $\WH{\mc L}^Ag=\mc Lg$,  
where $\WH{\mc L}^A$ is the 
infinitesimal generator of the semigroup 
$(\WH P^A_t)_{t\geq0}$.
\end{Lemma}
Lemma~\ref{lem:KeyLemma} 
is a straightforward consequence of the definition of
the infinitesimal generator and 
the Hille-Yosida theorem, see e.g. \cite[Lemma 31.7]{Sato}
(see~\cite{MP_2010_LongVersion} for the complete proof of Lemma~\ref{lem:KeyLemma}).
If $\WT S^A$ and $\WH S^A$ are themselves Feller processes, 
the relations between $\WT P^A$ and $\WT{\mathcal L}^A$, 
and between $\WH P^A$ and $\WH{\mathcal L}^A$ 
can formally be expressed as follows:
\begin{equation}\label{eq:PA}
\WT P^A_t = \exp\le(t\WT{\mathcal L}^A\ri),
\qquad\qquad
\WH P^A_t = \exp\le(t\WH{\mathcal L}^A\ri).
\end{equation}
Equation \eqref{eq:PA} can be given a precise meaning if, for example, 
$\WT P^A_t$ and $\WH P^A_t$ 
can be defined as a self-adjoint operator on a 
separable Hilbert space (see e.g. Ch. XII in Dunford and Schwarz \cite{DunfordSchwarz},
or Hille and Philips \cite{HillePhilips}). By determining the spectral 
decompositions of $\WT{\mathcal L}^A$ and $\WH{\mc L}^A$ one can construct spectral expansions of $\WT P_t^Af(x)$ and $\WH P_t^Af(x)$,
which in the case of a discrete spectrum reduces to a series expansion.
See Linetsky \cite{Linetskylookback,LinetskyIJTAF,LinetskyAsian} for a development of this 
spectral expansion approach for one-dimensional diffusion models in finance, and an overview of related literature.

When (asymmetric) jumps are present, the operator is non-local and 
not self-adjoint, and the spectral
theory has been less well developed, with fewer explicit
results. Here we will follow a different
approach: we will approximate $S$
by a finite state Markov chain, and show that for the approximating chain a matrix analog of the identities \eqref{eq:LA}--\eqref{eq:PA} holds true,
where the infinitesimal generators $\WT{\mathcal L}^A$ and $\WH{\mathcal L}^A$
can be easily obtained from $\mathcal L$.
We give a self-contained development of this approach 
in Section \ref{sec:Barriers_MC}.

%% file: barriersforchains1.tex
Given a Markov price process $S$ of interest, the idea is to construct 
a continuous-time finite state Markov chain $X$ whose dynamics are ``close''
to those of $S$, and to calculate the relevant expectations for 
this approximating chain. 
In this section we will focus on the latter; we will return to 
the question of how to construct such a  
chain in Section \ref{sec:MarkovChainConstr}. 
Assume therefore we are given a finite state 
continuous-time 
Markov chain $X=\{X_t\}_{t\geq0}$. 
From Markov chain theory it is well known that the chain  
is completely specified by its state-space 
(or grid) $\mathbb G\subset\E$
and its generator matrix $\Lambda$, 
which is an $N\times N$ square matrix with zero row sums and 
non-positive diagonal elements, if $\mathbb G$ 
has $N$ elements.
Given the generator matrix $\Lambda$ 
the family of transition matrices
$(P_t)_{t\geq0}$ of $X$,
defined by
$P_t(x,y):=\P_x[X_t=y]$
for
$x,y\in \mathbb G$,
is given  by
$$P_t=\exp(t\L).$$
In particular, the expexted 
discounted pay-off
$\phi(X_T)$
at maturity $T$ is then given by
\begin{eqnarray}
\label{eq:SemiGp}
\EE_x[\te{-rT}\phi(X_T)] &= &
\te{-r T} \cdot \left(\exp(T\L)\phi\right)(x)
\end{eqnarray}
for $x\in\GG$ and any function $\phi:\mathbb G\to\RR$. 
Here and throughout the paper we will identify any square matrix
$\A\in\RR^{N\times N}$ and any vector $\phi$ in $\RR^N$ with
functions
\begin{eqnarray*}
\A:\mathbb G\times \mathbb G\to \RR,& & \qquad\A(x,y):=e_x'\A e_y,\qquad x,y\in \mathbb G,\quad\text{and}\\
\phi:\mathbb G\to \RR,& & \qquad\phi(x):=e_x'\phi, \qquad x\in \mathbb G,
\end{eqnarray*}
where
the vectors
$e_x,e_y$
denote the corresponding standard basis vectors
of
$\RR^N$
and
$'$
stands for transposition. 

The generator $\Lambda$ can be retrieved from 
the family of matrices 
$(P_t)_{t\ge0}$ 
defined above
by differentiation at $t=0$, that is,
\begin{equation*}
P_t = I + t \Lambda + o(t) \quad
\text{ as }\> t\searrow 0. 
\end{equation*}
In order for $X$ itself to form a pricing model, 
defined under a martingale measure, we will suppose in addition 
that $X$ is non-negative and that the discounted process is a martingale; more precisely we assume that  $\{M_t\}_{t\ge 0}$ is a martingale where
\begin{equation}\label{eq:mart}
M_t = \begin{cases}
\te{-\gamma t}X_t,\, & t<\zeta,\\
\te{-\gamma \zeta}X_\zeta, & t\ge\zeta,
\end{cases}
\end{equation}
with $\zeta=\inf\{t\ge0: X_t\in\partial\GG\}$ the hitting time of the 
``boundary'' $\partial\GG$ which consists 
of the smallest and largest elements of the state space 
$\GG$, and $\gamma=r-d$. 
The Markov property of $X$ implies that 
the process $M$ given in \eqref{eq:mart} is a martingale precisely if
\begin{equation}\label{eq:meta}
(\Lambda \eta)(x) = \gamma\, \eta(x)\q\q\q
\text{for $x\in \GG\backslash\partial\GG$},
\end{equation}
where the function $\eta:\GG\to\mbb E$ is given by $\eta(z) = z$.
Below
we will show how to express the exit probabilities of the chain using matrix
exponentiation in a way that is  identical in form to the expected value~\eqref{eq:SemiGp}
of a European pay-off.
To that end, we partition $\mathbb G$ into a `continuation' set
 $\WH{\mathbb G}$ and a `knock-out' set 
 $\WH{\mathbb G}^c := \mathbb G\backslash\WH{\mathbb G}$, where
\begin{eqnarray}
\label{eq:E_Hat} \WH{\mathbb G} & := & \{x\in \mathbb G\>: x\in A^c\},
\end{eqnarray}
 and define the first exit time of $X$ 
from $\WH{\mathbb G}$ by
\begin{equation}\label{eq:tau}
\tau := 
\inf\{t\in\RR_+\>:\>X_t\notin\WH{\mathbb G}\},
\end{equation}
where we use the convention $\inf\emptyset: = \infty$ and where 
we will take the set $A$ as in \eqref{eq:A}.

The value of a general barrier knock-out option with a rebate
depends on the joint distribution of  
the exit time $\tau_A$ from $A$ and the positions of the underlying
at maturity and at the moment of exit.
The corresponding quantities for the chain $X$
can be expressed in terms of two 
transformations of
$X$,
namely the 
chain $\WH X$ that is killed upon exiting $\WH{\mathbb G}$ and 
the chain $\WT X$ 
that is absorbed at that instance, respectively. Correspondingly, 
we associate to the generator matrix $\Lambda$, two matrices: 
the $\WH N\times\WH N$ matrix $\WH\Lambda$, 
where $\WH N:=|\WH{\mathbb G}|$, and the $N\times N$ matrix $\WT\Lambda_r$,
defined by
\begin{eqnarray}
\label{eq:Def_tilA}
\WT\Lambda_r(x,y) & := &  
\begin{cases} \Lambda(x,y) - r &\text{if $x\in\WH{\mathbb G}$, $x=y$,}\\
\Lambda(x,y) & \text{if $x\in\WH{\mathbb G}$, $y\in\mathbb G$, $x\neq y$,}\\
0 & \text{if $x\in \WH{\mathbb G}^c$, $y\in \mathbb G$,}
\end{cases}\\
\label{eq:Def_tA}
\WH\Lambda(x,y) & := & \WH\Lambda_0(x,y)  :=  
\Lambda(x,y) \qquad \text{if $x\in\WH{\mathbb G}$, $y\in\WH{\mathbb G}$.}
\end{eqnarray}
We can now state the key result of this section:

\begin{Thm}
\label{thm:Main} For any $T>0$, $x\in \mbb G$ and $r\ge 0$ 
and any function $\phi:\mbb G\to\mathbb R$ it holds that
\begin{equation}\label{eq:main}
\EE_x\left[\te{-r(T\wedge\tau)}\phi(X_{T\wedge\tau})\right]
= \left(\exp\left(T\WT\L_r\right)\phi\right)(x).
\end{equation}
In particular, for $\psi:\WH{\mbb G}\to\mathbb R$ and 
$\xi: \mbb G\to\mathbb R$ with $\xi(x)=0$ for $x\in\WH{\mbb G}$
we have that
\begin{eqnarray}
\label{eq:BarrierPrice} \EE_x\!\!\left[\psi(X_T)\I_{\{\tau>T\}}\right]\!\! &=&
\!\!\left(\exp\left(T\tL\right)\psi\right)(x)\qquad\q\text{for any} \quad
x\in\WH{\mbb G},\\
\label{eq:BarrierPrice3}
\EE_x\!\!\left[\phi(X_T)\I_{\{\tau \leq  T\}}\right]\!\!
&=&\!\! \le(\exp\le(\Lambda T\ri)\phi\ri)(x) - 
\left(\exp\left(T\WH\L\right)\WH\phi\right)(x)\,\I_{\WH{\mbb G}}(x) \ \ \text{for any} \quad
x\in \mbb G,\\
\label{eq:BarrierPrice2}
\EE_x\!\!\left[\te{-r\tau}\xi(X_\tau)\I_{\{\tau \leq  T\}}\right]\!\!
&=& \!\!\left(\exp\left(T\WT\L_r\right)\xi\right)(x)\qquad\q\text{for any} \quad
x\in \mbb G,
\end{eqnarray}
where $\WH\phi=\phi|_{\WH{\mbb G}}$, 
the restriction of $\phi$ to $\WH{\mbb G}$.
\end{Thm}

%
%

Formulas~\eqref{eq:BarrierPrice}--\eqref{eq:BarrierPrice2} give 
a simple way of computing barrier option prices by a single matrix
exponentiation.
The expectation in~\eqref{eq:BarrierPrice}
can be obtained by computing the spectral decomposition of the matrix
$\tL=UD U^{-1}$
and applying the formula
$\exp\left(T\tL\right)=U\exp(TD) U^{-1}$.
%
%
The powerful Pad\'e approximation method  
for matrix exponentiation, 
described in~\cite{Higham}, can 
also be used
to compute efficiently the matrix exponentials 
in Theorem~\ref{thm:Main}.
Since the state-space is finite,
Theorem~\ref{thm:Main} is a corollary 
of Lemma \ref{lem:KeyLemma}. 
We present next a direct probabilistic derivation.

\begin{pr} 
To prove 
equation \eqref{eq:main}, we will verify that the expected value of an 
Arrow-Debreu barrier security
that pays 1 precisely if $X$ 
is in the state $y$ at the earlier of the maturity $T$ and the
knock-out time $\tau$ is given by 
\begin{equation}\label{eq:exit}
\mathbf E_x[\te{-r(T\wedge\tau)}\I_{\{X_{T\wedge\tau} = y\}}] = 
\le(\exp(T\WT\Lambda_r)\ri)(x,y) \qquad \text{for all $x,y\in{\mbb G}$}.
\end{equation}
For a given time grid 
$\mathbb T_n=(k\Delta t, k=0,1,2,\ldots,n)$ with $\Delta t = T/n$ 
denote by 
$\WT P^{(n)}_T(x,y)$ 
the expected value of the corresponding 
discretely monitored Arrow-Debreu security and let
$$\tau_n=\inf\{s\in\mathbb T_n: X_{s}\notin\WH{\mbb G}\}$$
be the corresponding time at which the barrier is crossed.
Since the paths of the chain $X$ are piecewise constant, it follows that 
$\tau_n\downarrow\tau$ and $X_{\tau_n}\to X_\tau$ as 
$n$ tends to infinity. Hence the expected values of the discretely monitored 
Arrow-Debreu securities 
converge to the expected value
of the continuously monitored one,  
$$\WT P^{(n)}_T(x,y)=\EE_x[\te{-r(T\wedge\tau_n)}\I_{\{X_{T\wedge\tau_n}=y\}}] 
\longrightarrow \mathbf E_x[\te{-r(T\wedge\tau)}\I_{\{X_{T\wedge\tau} = y\}}].$$
Clearly, since $\WH{\mathbb G}^c$ is the knock-out set, it holds for all $t\ge 0$ that
\begin{eqnarray*}
\WT P^{(n)}_t(x,y) &=& \begin{cases} 1 & \text{if $x\in\WH{\mathbb G}^c$, $x=y$}\\
0 & \text{if $x\in\WH{\mathbb G}^c$, $x\neq y$}\\
\end{cases}\\
&=& \le(\,I-\ovl I\,\ri)(x,y) \qquad\text{for all 
$x\in\WH{\mathbb G}^c$, $y\in\mathbb G$,}
\end{eqnarray*}
where $\ovl I$ is a square matrix of size $N$ with $\ovl I(x,x)=1$ if $x\in\WH{\mbb G}$ and zero else.
Further, for $x\in\WH{\mbb G}$ an application 
of the Markov property of $X$ shows that
\begin{eqnarray*}
\WT P^{(n)}_T(x,y) &=& \te{-r\Delta t}\sum_{z\in\mathbb G}P_{\Delta t}(x,z)\WT P^{(n)}_{T-\Delta t}(z,y)\\ 
&=& \le(\ovl I\,\le(\te{-r\Delta t}P_{\Delta t}\ri)%
\,\WT P^{(n)}_{T-\Delta t}\ri)(x,y). \end{eqnarray*}
Combining the two cases, iterating the argument 
and using the differentiability of $P_t$ at $t=0$ shows that 
\begin{eqnarray*}
\WT P^{(n)}_T(x,y) &=& \le(\le(I - \ovl I + \ovl I\,\te{-r\Delta t}P_{\Delta t}\ri)\WT P^{(n)}_{T-\Delta t}\ri) (x,y) \\
&=& \le(\le(I - \ovl I + \ovl I\,\te{-r\Delta t}P_{\Delta t}\ri)^{T/\Delta t}\ri)(x,y)\\
&=& \le(\le(I  + \ovl I\,\te{-r\Delta t}(P_{\Delta t}-I)+ \ovl I(\te{-r\Delta t} - 1)\ri)^{T/\Delta t}\ri)(x,y)\\
&=& \le(\le(I + \Delta t(\WT\Lambda_0 - r\ovl I)+o(\Delta t)\ri)^{T/\Delta t}\ri)(x,y), 
\end{eqnarray*}
since $\WT\Lambda_0=\ovl I \Lambda$. When $\Delta t=T/n$ tends to zero, 
this expression converges to $\le(\exp\le(T\WT\Lambda_r\ri)\ri)(x,y)$, which completes the proof of \eqref{eq:exit}
and hence implies~\eqref{eq:main}.
Equation~\eqref{eq:BarrierPrice2}  
follows then directly 
by applying~\eqref{eq:main} to $\xi$. 
Noting that for any $\psi:\WH{\mathbb G}\to\mathbb R$
and any
$x\in\WH{\mathbb G}$
we have
$$\le(\WH{\Lambda} \psi\ri)(x) = \le(\WT{\Lambda}_0\psi_0\ri)(x),\qquad\text{where 
$\psi_0:\mathbb G\to\mathbb R$ is given by $\psi_0(y):=
\psi(y)\I_{\WH{\mathbb G}}(y)$},$$ 
we get that 
$$\le(\exp(T\WH{\Lambda})\psi\ri)(x)=\le(\exp(T\WT{\Lambda}_0)\psi_0\ri)(x),$$
which yields~\eqref{eq:BarrierPrice}.  Finally, \eqref{eq:BarrierPrice3} is a direct consequence of \eqref{eq:SemiGp}, \eqref{eq:BarrierPrice} 
and the fact that a European option with pay-off $\phi(X_T)$ 
is equal to the sum of a knock-out barrier option and a 
knock-in barrier option with the same pay-off $\phi(X_T)$ 
and same knock-out/knock-in levels. 
\end{pr}

%% file: construction.tex
\begin{figure}[t]
\includegraphics[width=\textwidth]{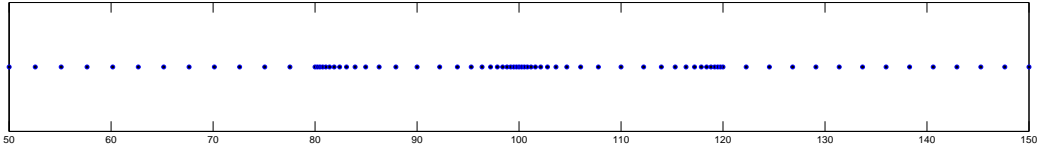}
\caption{\footnotesize{ 
The figure shows a non-uniform grid that was generated using the algorithm described in Section \ref{sec:MarkovChainConstr}. 
The grid consists of 
$N=75$ points that are densely distributed 
around the spot $S_0=100$ and 
the upper and lower barriers $\ell=80$ and $u=120$; 
its smallest and largest values were chosen as 
$x_1=50$ and $x_N=150$, and $N_i=25$, $i=1,2,3$, points 
were placed in the areas $[x_1,\ell]$, $[\ell,u]$ and $[u,x_N]$ 
with respective density parameters
$d_1^-=100$, $d_1^+=1$, $d_2^-=1$, $d_2^+=1$, and $d_3^-=1$, 
$d_3^+= 100$.
}}
\label{fig:StateSpace}
\end{figure}

The Markov chain approximation algorithm for homogeneous 
Feller processes can now be described as
follows:
\begin{enumerate}
\item[(1)] Construct an approximating Markov chain:
\begin{enumerate}
\item[(a)] specify a (non-uniform) grid 
$\GG$;
\item[(b)] define a generator matrix 
$\L$
of a Markov chain with state-space 
$\GG$.
\end{enumerate}
\item[(2)] Compute the barrier option prices using 
formulae~\eqref{eq:main}--\eqref{eq:BarrierPrice2}.
\end{enumerate}

A suitable choice of the grid in step 
(1a) is essential for the effectiveness of the 
above pricing algorithm.
The construction of an optimal grid (according to some criterion)
is a topic of separate study, which will not be pursued further in this 
paper.
One of the features of a good grid is that it has
sufficient resolution in regions of 
interest, such as the current spot value and the barrier levels, 
which is a necessary condition for constructing 
a Markov chain market model 
that
approximates well
the dynamics of the given price process.
Another desirable feature is that the grid ``covers'' a sufficiently 
large part of the state space, which is needed to control the 
truncation error that arises when approximating an infinite 
state space by a finite state space. To employ
a uniform grid that satisfies these conditions 
would be computationally expensive.
The use of adaptive meshes for option pricing 
was proposed in Figlewski and Gao~\cite{Gao_Figlewski}.
Here we employ the following 
procedure  
for generating 
a suitable non-uniform grid 
$\GG$, 
based on an algorithm from~\cite{TavellaRandall}:

\bigskip
\begin{enumerate}
\item Pick 
$N_i\in\NN$
and the density parameters
$d_i^\pm\in(0,\infty)$,
$i=1,2,3$,
and
the smallest and largest values 
$x_1, x_N$
of the grid
$\GG$,
where
$N=N_1+N_2+N_3$.
\item Define 
$\GG_i:=$\textsc{GenerateSubGrid}$(a_i,s_i,b_i,N_i,d_i^-,d_i^+)$,
for 
$i=1,2,3$,
where
$a_1=x_1, s_1= \ell, b_1=(S_0+\ell)/2=a_2, s_2 = S_0, b_2=(u+S_0)/2=a_3, s_3=u,
b_3=x_N$.
\item  Set $\GG:=\GG_1\cup\GG_2\cup\GG_3$,
\end{enumerate}

\noindent where the subgrid is generated by the following procedure:
\medskip

\noindent \textsc{GenerateSubGrid}$(a,s,b,M,g_1,g_2)$
\begin{enumerate}
\item Compute $c_1 = \mathrm{arcsinh}\left(\frac{a-s}{g_1} \right)$,
$c_2 = \mathrm{arcsinh}\left(\frac{b-s}{g_2} \right)$.
\item Define the lower part of the grid by the formula
$x_k := s + g_1 \mathrm{sinh}(c_1(1-(k-1)/(M/2-1)))$
for
$k\in\{1,\ldots,M/2\}$.
Note that $x_1=a, x_{M/2}=s$.
\item Define the upper part of the grid using the formula
$x_{k+M/2} := s + g_2 \mathrm{sinh}(c_2 2k/M)$
for
$k\in\{1,\ldots,M/2\}$.
Note that
$ x_{M}=b$.
\end{enumerate}
\textsc{Return} $\{x_1,\ldots,x_M\}$
\bigskip

The
non-uniform state-space
$\GG\subset\E$
for the Markov chain
$X$
is constructed by concatenating the three subgrids 
that are generated by specifying lower, middle and
upper points
$a<s<b$
and density parameters
$g_1$
and
$g_2$.
A smaller density corresponds to a grid that
is more concentrated around the middle point 
$s$.
Observe that the algorithm above places the current spot
$S_0$
and the barrier levels
$\ell<u$
on the grid
$\GG$
and that the resolution of the grid around 
$\ell, S_0$
and
$u$
can be controlled by the density parameters
$d_i^\pm$,$i=1,2,3$.
A Matlab implementation of this grid generator 
can be downloaded from \cite{MG_Code}. 
The remainder of this
section will be devoted to step (1b)
in the algorithm described 
above.

\subsection{Jump processes with state-dependent characteristics}
\label{sec:StateDepChar}
The construction of the generator matrix 
$\L$
of the approximating Markov 
chain 
$X$
is now carried out in two steps: we first define the jump matrix
$\L_J$,
which corresponds to the discretization of the jump measure
$\nu$,
and then characterize a tri-diagonal generator matrix
$\L_D$
by stipulating that the Markov chain 
$X$
with the generator
$\Lambda= \L_J+\L_D$
has the same instantaneous moments as the process
$S$.

We start by building the state-space
$\GG\subset\E$
with 
$N$
elements
using the algorithm described in the beginnig of
this section.
Define the sets
\begin{eqnarray*}
\partial \GG:=\{x_1,x_N\}\qquad\text{and}\qquad 
\GG^o:= \GG\backslash\partial \GG,
\end{eqnarray*}
where the ``boundary''
$\partial \GG$
consist of the smallest
(i.e. $x_1$)
and largest 
(i.e. $x_N$)
elements
in
$\GG$
and the ``interior''
$\GG^o$
is the complement of the boundary.
For any given 
$x\in\GG^o$
we associate
to
$\GG$
the set 
$\GG_x\subset(-1,\infty)$
defined by
\begin{eqnarray*}
\GG_x & := & \left\{\frac{z}{x}-1\>:\>\> z\in\GG\right\}.
\end{eqnarray*}
The set 
$\GG_x$
consists of the relative jump sizes 
of jumps starting from
$x$
and arriving at any other point
in
$\GG$.
The 
$x$-th row of the
jump part 
$\L_J$
of the generator of 
$X$
is obtained by discretizing the jump measure 
$\nu(x,\td y)$
on the set 
$\GG_x$.
In particular let
$\GG_x=\{y_i\>:\>y_1<\ldots<y_{N-1}\}$
and define a function
$\alpha_x:\GG_x\to[-1,\infty]$
such that
\begin{eqnarray}
\label{eq:alphaX}
& & \alpha_x(y_i)\in(y_{i},y_{i+1})\quad\text{for}\quad i\in\{1,\ldots,N-1\}
\end{eqnarray}
where
$\alpha_x(y_0)=-1, \alpha_x(y_N)=\infty$.
A possible natural choice for
$\alpha_x(y_i)$
would be the mid-point of the interval
$$\alpha_x(y_i)=\frac{y_{i}+y_{i+1}}{2}.$$
We can now define the jump part of the generator as
\begin{eqnarray}
\label{eq:JumpGen}
\L_J(x,x(1+y_i)) & := & \int_{\alpha_x(y_{i-1})}^{\alpha_x(y_i)}\nu\left(x,\td y \right)
\end{eqnarray}
where 
$i\in\{1,\ldots,N\} \>\>\text{and}\>\> y_i\neq0$ 
and
\begin{eqnarray}
\L_J(x,x) & := & -\sum_{z\in \GG\backslash\{x\}}\L_J(x,z).
\label{eq:JumpGenIsGen}
\end{eqnarray}

Note that the function
$\alpha_x$
generates a partititon of 
$(-1,\infty)$.
The jump intensities of the chain, defined  in
formula~\eqref{eq:JumpGen},
are obtaiend by integrating the L\'evy measure 
over the corresponding part of the partition.
For 
$x\in\partial \GG$
we set 
$\L_J(x,y):=0$
for all
$y\in\GG$.
It is clear that the matrix
$\L_J$
constructed in this way 
is a generator matrix.

In the second step we 
match the first and second instantaneous
moments of the asset price process
$S$.
In other words 
the chain 
$X$
must satisfy conditions
\begin{eqnarray}
\label{eq:MomentMatchDiff}
\EE_{x}\left[(S_{\Delta t}-S_0)^j\right] & = &
\EE_{x}\left[(X_{\Delta t}-X_0)^j\right]+ o(\Delta t),\qquad\text{for}\quad x\in\GG^o,\>\>
j\in\{1,2\}
\end{eqnarray}
for all starting states
$x\in\GG^o$.
Note that condition~\eqref{eq:MomentMatchDiff} 
implicitly assumes that 
the second instantaneous 
moment of
$S$
exists. 
This is the case if
the jump measure satisfies
the following condition
\begin{eqnarray}
\label{eq:JumpSecondMomentCond}
\int_{(-1, \infty)} y^2\nu(x,\td y) <\infty \quad \text{for all}\quad x\in\E
\end{eqnarray}
which we now assume to hold. 

The task now is to find a tri-diagonal generator
matrix
$\L_D$
such that the chain generated by the sum
$\L_D+\L_J$
satisfes~\eqref{eq:MomentMatchDiff}.
The tri-diagonal matrix
$\L_D$
therefore has to satisfy the following conditions
\begin{eqnarray}
\label{eq:JumpGenEqu1}
\sum_{z\in \GG} \L_D(x,z) &=& 0\quad\text{and}\quad 
\L_D(x,z)+ \L_J(x,z)\geq0 \quad \forall z\in \GG\backslash \{x\},\\
\label{eq:JumpGenEqu2}
\sum_{z\in \GG} \L_D(x,z) (z-x) &=& (r-d) x - \sum_{z'\in \GG} \L_J(x,z') (z'-x), \\
\label{eq:JumpGenEqu3}
\sum_{z\in \GG} \L_D(x,z) (z-x)^2 &=&  x^2\left[\sigma\left(x\right)^2+\int_{-1}^\infty y^2\nu(x,\td y) \right]- \sum_{z'\in \GG} \L_J(x,z') (z'-x)^2
\end{eqnarray}
for all
$x\in\GG^o$,
where
$r,d$
are the instantaneous interest rate and dividend yield respectively
and 
$\sigma$
is the local volatility function in~\eqref{eq:gengen}.
The right-hand side of equation~\eqref{eq:JumpGenEqu2}
is the difference of the risk-neutral drift and the drift 
induced by the presence of jumps.
Similarly the right-hand side of the linear equation in~\eqref{eq:JumpGenEqu3}
consists of the difference of the
instantaneous second moments of the asset price process 
$S$
(computed directly from its generator~\eqref{eq:gengen})
and the chain that corresponds to the jump 
generator
$\L_J$.
As usual we assume the absorbing boundary condition 
$\L_D(x,y)=0$
for all
$x\in\partial\GG,\>y\in\GG$.

The linear system
in~\eqref{eq:JumpGenEqu1}--\eqref{eq:JumpGenEqu3}
can typically be satisfied by 
a tri-diagonal generator matrix 
$\L_D$
if 
$\sigma(x)$
is strictly positive.
Once we find
$\L_D$
we define the generator matrix of the approximating chain
$X$
by
$$
\L:=\L_D+\L_J.
$$

\begin{Rem}\rm
\label{rem:PhillipsApp}
In the case that 
$S$
is a diffusion time-changed by an independent L\'evy subordinator,
there is an alternative approach to constructing the approximating
continuous-time Markov chain $X$, based on the Phillips theorem.
For the details of this construction see~\cite{MP_2010_LongVersion}.
\end{Rem}




%% file: convergence.tex
\subsection{Convergence of barrier option prices}
Consider a sequence
of finite-state continuous-time
Markov chains 
$X^{(n)}$ 
approximating 
a given Feller price process 
$S=\{S_t\}_{t\ge 0}$.
For $X^{(n)}$ to replicate as closely as possible the dynamics of
$S$ one chooses the generator matrix $\Lambda^{(n)}$ with the 
corresponding state-space $\GG^{(n)}$ such that it is uniformly close to the 
infinitesimal generator $\mc L$ of $S$, in the sense that 
the distance $\epsilon_n(f)$ between the generators
is small for a sufficiently large class $\WT{\mathcal D}$
of regular test functions $f$, where
\begin{equation*}
\epsilon_n(f) := \max_{x\in(\GG^{(n)})^o}\le|\Lambda^{(n)} f_n(x) - \mathcal L f(x)\ri|,
\end{equation*}
where
$(\GG^{(n)})^o$
equals 
$\GG^{(n)}$
without the smallest and the largest elements
and 
$f_n=f|_{(\mathbb G^{(n)})^o}$ 
is the restriction of $f$ to $(\mathbb G^{(n)})^o$.
More specifically, if $\epsilon_n(f)$ tends to zero as 
$n$ 
tends to infinity for 
$f$ 
in the class 
$\WT{\mathcal D}$
and the probability that the chain exits $(\mathbb G^{(n)})^o$ before time $T$ tends to zero, 
then the sequence of processes $(X^{(n)})_{n\in\NN}$ converges 
weakly to the process $S$. 
This weak convergence on the level of the process implies in particular that the marginal distributions 
of $X^{(n)}$ will converge to those of $S$, and therefore 
the values of European options converge, that is,
\begin{equation*}
\EE_x[f(X^{(n)}_T)] \to \EE_x[f(S_T)]
\end{equation*}
for $x\in\E$, maturity $T>0$, and continuous bounded functions $f$. 

The payoff of the barrier option can be described in terms of 
the first-passage time of $S$ and the position of $S$ at that 
moment, which are both functionals of the path $\{S_t\}_{t\ge 0}$.
For the weak convergence of $X^{(n)}$ to $S$ to carry 
over to convergence of barrier-type payoffs, 
continuity (in the Skorokhod topology) is required of 
these two functionals, which is guaranteed to hold under 
Assumption \ref{as1}. 
In view of the fact that the payoff of 
a barrier option is typically a discontinuous function, 
an additional condition is needed
to ensure the convergence of the barrier option prices; we will 
assume that $\ell$ and $u$ are such that
\begin{equation}\label{eq:XT}
\P_x\le(S_T  \in\{\ell,u\} \ri) = 0. 
\end{equation}
Most models used in mathematical finance satisfy this 
condition. 
Even if~\eqref{eq:XT} is not satisfied, 
this does not constitute a limitation in practice, 
since for any given process 
$S$ the condition is satisfied for all 
but countably many pairs $(\ell,u)$.  
The statement of the convergence is made precise in the following
theorem.

%

\begin{Thm}\label{thm:conv} 
Let $S$ be a Feller process with state-space $\E$ and 
infinitesimal generator $\mathcal L$
that does not vanish at zero and infinity.\footnote{The set
where the generator does not vanish is defined to be the set of all
$x\in\mathbb E\cup \{\infty\}$
with the following property: for every open interval 
$I$ 
in 
$\mathbb E$
that contains 
$x$
(if 
$x=\infty$
then 
$I$
takes the form
$(M,\infty)$
for some 
$M\in\mathbb E$)
there exists a function 
$f\in\mc D$
with compact support in
$I$
such that the function
$\mc L f$
is not identically equal to zero.}
Let $\left(X^{(n)}\right)_{n\in\mathbb N}$ be a sequence 
of Markov chains with generator matrices $\Lambda^{(n)}$ such that
the rows corresponding to the smallest and the largest elements in
$\GG^{(n)}$
are equal to zero. 
Assume further that 
the following two conditions are satisfied
for any function $f$ 
in a core of 
$\mathcal L$:\footnote{A core $\mathcal{C}$ 
of the operator
$\mc L$
is a subspace of the domain of 
$\mc L$
that 
is (i) dense in
$C_0(\E)$ 
and (ii) there exists
$\lambda>0$
such that the set 
$\{(\lambda - \mathcal L)f: f\in \mc C\}$ is dense in $C_0(\E)$.}
\begin{eqnarray}
& &
\label{eq:ConvHom}
\e_n(f)\to 0 \q\text{as $n\to\infty$},
\\
& & \text{either (i) $\lim_{y\searrow0}\mc Lf(y)=0$ or 
(ii) $\lim_{n\to\infty}\PP_x\left[\tau_{(\GG^{(n)})^o}^{(n)}>T\right]=1$},
\label{eq:CorrectCondition}
\end{eqnarray}
where for any set
$G\subset\mathbb E$
we define
$\tau_G^{(n)}=\inf\{t\ge0: X^{(n)}_t\notin G\}$.
If \eqref{eq:XT} holds, then, as $n\to\infty$,
\begin{eqnarray*}
\EE_x\le[
g\le(X^{(n)}_{T}\ri)\I_{\{\tau_A^{(n)}>T\}}\ri]
&\longrightarrow&
\EE_x\le[
g(S_{T})\I_{\{\tau_A>T\}}\ri],\\
\EE_x\le[\te{-r\tau_A^{(n)}}
h\le(X^{(n)}_{\tau^{(n)}_A}\ri)\I_{\{\tau_A^{(n)}\leq T\}}\ri]
&\longrightarrow&
\EE_x\le[\te{-r\tau_A} 
h(S_{\tau_A})\I_{\{\tau_A \leq T\}}\ri], 
\end{eqnarray*}
for any bounded continuous functions 
$g, h: \E\to\mathbb R$.
\end{Thm}

\begin{Rem}\rm
Since condition \eqref{eq:ConvHom} 
is required to hold for all 
$f$ in a core of $\mc L$, 
it follows that $\mathbb G^{(n)}$
will eventually `fill up' the part of the  state-space 
$\mathbb E$
where the generator does not vanish.
Indeed, if there were to exist an open interval that does not intersect 
$\mbb G^{(n)}$ for large $n$, then condition \eqref{eq:ConvHom} would not hold
for functions 
$f\in C^\infty(\mbb E)$ 
with compact support in this open interval satisfying
$\mc L f\neq 0$, as $\Lambda^{(n)}f_n\equiv0$. In particular,  
condition \eqref{eq:ConvHom}
implies that $x^{(n)}_1\to 0$ and $x^{(n)}_n\to\infty$ 
when 
$\mathbb G^{(n)} = \{x^{(n)}_1, \ldots, x_n^{(n)}\}$ 
with 
$x^{(n)}_1< \ldots < x_n^{(n)}$
if the generator does not vanish in a neighbourhood of zero and infinity.
\end{Rem}

\begin{Rem}\rm
\label{Rem:Important}
Note that, since 
$\mc L f$
is by definition an element in
$C_0(\mathbb E)$
for any
$f$
in the domain of
$\mc L$,
it holds that 
$\mc Lf(y)\to0$
as 
$y\nearrow\infty$
and that 
$\mc Lf(0)$
is well defined and equal to
$\lim_{y\searrow0}\mc Lf(y)$.
Assume that 
$\mc L$
does not vanish at zero and infinity and that 
the rows corresponding to the smallest and the largest elements in
$\GG^{(n)}$
of the generator matrices $\Lambda^{(n)}$ 
are equal to zero.
Then, if conditions~\eqref{eq:ConvHom} and~\eqref{eq:CorrectCondition}(i)
are satisfied, 
we have
\begin{equation}\label{eq:GoodConv}
\max_{x\in\GG^{(n)}}\le|\Lambda^{(n)} f_n(x) - \mathcal L f(x)\ri|\to0\qquad\text{as}\quad
n\to\infty.
\end{equation}
\end{Rem}

\begin{Rem}\rm In practice the condition of boundedness of the 
payoff $f$ is not restrictive as it is always possible to 
consider the truncation $f\wedge M$ for large 
constants $M$ 
without losing noticeable accuracy. 
Further, under additional regularity properties on the 
parameters of the 
process $S$, the convergence in Theorem \ref{thm:conv} also holds true 
for barrier call options. To see why this is the case note that 
\begin{equation}\label{eq:cm}
\EE_x\le[(S_T - K)^+\I_{\{\tau_A<T\}}\ri] = x\, \ovl\EE_x\le[(1 - K/S_T)^+\I_{\{\tau_A<T\}}\ri]
\end{equation}
where $\ovl{\EE}$ denotes the expectation under the measure 
$\ovl{\mbf P}$ 
given by $\td\ovl{\mbf P}_x|_{\mathcal F_t}=x^{-1}S_t\td\mbf P_x|_{\mathcal F_t}$. 
Under $\ovl{\mbf P}$ the process 
$S$ remains a Markov process, and, under additional
regularity properties
$S$ is still a Feller process. The convergence then follows 
from Theorem \ref{thm:conv} as the pay-off function on 
the right-hand side of \eqref{eq:cm} is bounded.
\end{Rem}

\begin{Rem}\rm
Theorem~\ref{thm:conv}
follows by combining Ethier and Kurtz \cite[Theorem 4.2.11]{EthierKurtz}
and results in Section~VI.2 in Jacod and Shiryaev~\cite{JacodShiryaev}.
The complete proof of Theorem~\ref{thm:conv} can be found in~\cite{MP_2010_LongVersion}.
\end{Rem}

\subsection{Error estimates}\label{}
In this section we quantify the speed at which the 
algorithm converges by providing error estimates 
for a specific choice 
of a sequence of approximating Markov chains, assuming 
sufficient smoothness of the value-function of the barrier option.
Consider a Feller process $S$ 
with infinitesimal generator acting on $v\in C^{2}(\E)$ 
with compact support as %
\begin{eqnarray}
\mathcal L v(x) &=&  \mc L_{D}v(x) + \mc L_{J}v(x)\qquad\text{with}
\label{eq:gengengen}
\\
\mc L_{D}v(x)&:=& \frac{\sigma^2(x)x^2}{2}v''(x) + \gamma x v'(x),
\label{eq:L_Diff} 
\\ 
\mc L_{J}v(x)&:=&\int_{(-1,\infty)}[v(x(1+y)) - v(x) - v'(x)xy]
g(x,y) \td y,
\label{eq:L_Jump}
\end{eqnarray}
where $'$ denotes differentiation with respect to $x$, 
$\gamma=r-d$
($r$
and
$d$
are the instantaneous interest rate
and dividend yield respectively)
and $g:\mathbb E\times\left((-1,\infty)\backslash\{0\}\right)\to\mbb R_+$ is 
a nonnegative locally bounded function
such that 
$$
\int_{(-1,\infty)\backslash\{0\}} g(x,y)(|y|\wedge y^2)\td y < \infty\qquad \text{for all $x\in\mbb E$},
$$
where $x\wedge y = \min\{x,y\}$.

We next describe 
the sub-class of Markov processes that we will consider.

\begin{Def}\rm The process $S$ is called {\em uniformly of bounded 
jump-variation} if the jump-density $g$ satisfies the integrability condition
\begin{equation}
C_0 := \sup_{x\in[\ell,u]} \mc I(x)  < \infty, 
\end{equation}
where
\begin{equation}\label{Itx}
\mc I(x):= \int_{(-1,1)\backslash\{0\}} |y|g(x,y)\td y.
\end{equation}
\end{Def}

\begin{Def}\rm
The process $S$ is called {\em locally of stable type} on $[\ell,u]$ if there exist constants $\ovl\kappa_\pm, \unl\kappa_\pm > 0$ 
and $\alpha_\pm\in(0,2)$  such that for all 
$x\in[\ell,u]$
\begin{eqnarray}\label{stable-up}
\frac{\unl\kappa_+}{|y|^{1 +\alpha_+}} &\leq& g(x,y) \leq \frac{\ovl\kappa_+}{|y|^{1 +\alpha_+}},\qquad y\in(0,1),\\
\frac{\unl\kappa_-}{|y|^{1 +\alpha_-}} &\leq& g(x,y) \leq \frac{\ovl\kappa_-}{|y|^{1 +\alpha_-}},\qquad y\in(-1,0).
\label{stable-down}
\end{eqnarray}
\end{Def}
We will consider three different cases:
\begin{itemize}
\item \textbf{Case O}: $S$ is uniformly of bounded jump-variation.
\item \textbf{Case I}: $S$ is locally 
of stable type with $\alpha_-=1$ or $\alpha_+=1$.
\item \textbf{Case II}: $S$ is locally of stable type with $\alpha_\pm\in(1,2)$.
\end{itemize}
Case I can be considered to be a boundary case separating cases O and II,
 as $S$ is uniformly of bounded jump-variation if it is of stable type with $\alpha_\pm<1$.
The class of processes of uniformly bounded jump-variation (case O) 
contains processes $S$ whose jump-part forms a compound Poisson process 
(such as the Kou~ model \cite{Kou}), as well as the  
L\'{e}vy models of bounded variation 
such as for example the VG model. Examples of processes 
satisfying the condition in cases I and II are the Generalised 
Hyperbolic L\'{e}vy models and CGMY models with $Y>1$, respectively.
More generally, the class of processes that are locally of stable 
type contains the class of {\em regular 
L\'{e}vy processes of exponential type} studied in Boyarchenko 
and Levendorskii \cite{TheBook}.

We will impose in addition the following regularity conditions:
\begin{As}\label{as:bounded}
There exists a constant $C'$ such that for all $x\in[\ell,u]$
$$\sigma^2(x)x^2 + |\gamma|x + \int_{(-1,\infty)\backslash\{0\}} (y^2\wedge |y|) g(x,y)\td y  \leq C'.$$ 
\end{As}
\begin{As}\label{as:f}
The pay-off function $f$ is Lipschitz-continuous\footnote{A function $f:\E\to\mathbb R$ is Lipschitz continuous if there exists a constant $L>0$ such that $|f(x)-f(x)|\leq L|x-y|$ for all $x,y\in\E$.} and 
has compact support. The barrier option value-function 
$\WT P^A_tf(x)=\EE_{x}[\te{-r(t\wedge\tau_A)}f(S_{t\wedge\tau_A})]$ is $C^{1,3}([0,T]\times\E)$. \end{As}
\begin{Rem}\rm
In the strictly elliptic case ($\sigma(x)>0$ for $x\in[\ell,u]$) with a jump-part that is uniformly of bounded variation, explicit sufficient 
conditions on the data $f,g$ and $\sigma^2$ 
guaranteeing the smoothness of $v_A:(t,x)\mapsto \WT P^A_tf(x)$ 
required in Assumption \ref{as:f} 
follow from classical existence and uniqueness results for Cauchy-Dirichlet problems associated to second order partial-integro differential operators. Specifically, Theorem II.3.3 in Garroni \& Menaldi \cite{GarroniMenaldi} implies that $v_A\in C^{1,3}([0,T]\times [\ell,u])$ if, for some $0<\epsilon<1$,
the following conditions are satisfied: 
(a) $f$ has compact support in $(\ell,u)$ and $f\in C^{2+\epsilon}([\ell,u])$%
\footnote{$C^{n+\epsilon}([\ell,u])$ is the set of functions 
$f: [\ell,u]\to\mathbb R$ that are $n$ times continuously differentiable with the $n$th derivative $\epsilon$-H\"{o}lder continuous.}, (b) $\sigma^2\in C^{1}([\ell,u])$ and
(c) $x\mapsto g(x,y)|y|^{2+\epsilon}$ is 
$C^1([\ell,u])$ with bounded derivatives, uniformly 
over $(x,y)\in[\ell,u]\times(-1,\infty)$. 
Note that (c) implies that the $\mc L_Jv$ in \eqref{eq:L_Jump} 
is $C^1([\ell,u])$ if $v$ is $C^2([\ell,u])$.
\end{Rem}
Throughout the rest of this section we take $$0 < \ell < u < \infty$$ and 
consider the spatial grid given by
\begin{equation}\label{eq:GT}
\begin{aligned}
\GG^{(n)} &= \{x_i^{(n)}:0= x_1^{(n)} < \ldots < x_n^{(n)}\}.
\end{aligned}
\end{equation}
We will denote by 
\begin{equation}\label{eq:hd}
h(n)=\max_i\le|x^{(n)}_{i+1}-x^{(n)}_i\ri|
\end{equation}
the mesh of the spatial grid, and by
\begin{equation}\label{eq:k}
k(n) = \max_{x^{(n)}_k\in\WH{\mbb G}^{(n)}}
\int_{(-1,\infty)\backslash[L(x^{(n)}_k),U(x^{(n)}_k)]}
g(x^{(n)}_k,y)\td y,
\end{equation}
the tail mass of the jump-measure, 
where $\WH{\mbb G}^{(n)} = \mbb G^{(n)}\cap[\ell,u]$ and 
$$L(x^{(n)}_k):= \frac{x^{(n)}_2}{x^{(n)}_k} - 1\q
\text{and} \q U(x^{(n)}_k) := \frac{x^{(n)}_{n-1}}{x^{(n)}_k} - 1$$ 
are the relative jump-sizes from $x^{(n)}_k$ 
to the one-but-largest and one-but-smallest elements $x^{(n)}_2$ and $x^{(n)}_{n-1}$ of the state-space $\mbb G^{(n)}$.  We approximate $S$ by 
a sequence of 
Markov chains $(X^{(n)})_{n\in\mathbb N}$ 
with infinitesimal generators 
$\Lambda^{(n)}$.
An explicit description of
$\Lambda^{(n)}$ 
is given in Appendix~\ref{a:MC}.

\begin{Thm}\label{prop:error}
Let Assumptions  \ref{as:bounded} 
and \ref{as:f} hold, and consider cases O, I or II. 
Assume also that the sequences of grids 
$\mbb G^{(n)}$ 
satisfies 
$$
\lim_{n\to\infty} h(n) = \lim_{n\to\infty} k(n) =  0.
$$
Then there exist constants $C_1$, $C_2$,  
independent of $h(n)$ and $k(n)$, given in \eqref{eq:hd} 
and 
\eqref{eq:k}, 
such that for all $n$ 
sufficiently large and $x\in\mathbb G^{(n)}$

\begin{eqnarray*}
\le|\EE_{0,x}\le[\te{-\int_0^{T\wedge\tau_A^{(n)}}r^{(n)}(t)\td t}f\le(X^{(n)}_{T\wedge\tau^{(n)}_A}\ri)\ri] - \EE_{0,x}\le[\te{-\int_0^{T\wedge\tau_A}r(t)\td t}f\le(S_{T\wedge\tau_A}\ri)\ri]
\ri|&\leq& C_1 E(h) + C_2 k,\\
&\phantom{=}&
\end{eqnarray*}
where $h=h(n)$, $k=k(n)$ and
\begin{equation}\label{E}
E(h) = \begin{cases}h, & \text{in cases O and II},\\ 
- h\log h, & \text{in case I}.\\
\end{cases}
\end{equation}
\end{Thm}
The proof of Theorem \ref{prop:error} is given in  
Appendix \ref{a:error}. The logarithmic factor in the error bound \eqref{E}
in case I is due to the form \eqref{stable-up}--\eqref{stable-down} 
of the singularity of the jump measure at zero in this case (see the proof 
for details). Note that the discontinuous part of a process satisfying 
case I is of infinite variation, but has zero $(1+\epsilon)$-variation for every $\e>0$. 

Numerical experiments, 
reported in Section \ref{sec:numericalEx}, suggest that 
in several pricing models of interest
the error of the Markov generator (MG) algorithm described in Section 
\ref{sec:MarkovChainConstr} is actually of order $h^2$. 
A further theoretical investigation of error bounds and 
sharp rates of convergence of the algorithm, 
is left for future research.

%% file: numericalex1.tex
In this section the Markov chain algorithm
is examined numerically in a variety of contexts. 
Subsection~\ref{subsub:GY} contains 
a numerical examples for the geometric Brownian motion
model
and a comparison with a binomial tree method 
for the pricing of barrier options. 
In subsection~\ref{subsec:Levy} 
the algorithm is applied to two cases of 
L\'evy driven SDEs. Further numerical examples, 
including comparisons of computational times
of the various algorithms, can be found 
in~\cite{MP_2010_LongVersion}.


\subsection{Geometric Brownian motion}
\label{subsub:GY}

The model is given by SDE~\eqref{eq:dif}
where the volatility function 
$\sigma(x)=\sigma_0$
is constant and the drift equals
$\gamma = r-d$,
where 
$r$
is the risk free rate and 
$d$
is the dividend yield.
We now compare our algorithm (MG), based on the Markov generator 
of the approximating chain 
$X$,
with the results obtained in Geman and Yor~\cite{GemanYor}
and Kunitomo and Ikeda~\cite{KunitomoIkeda}.
The numerical results are contained in 
Table~\ref{t:BS_Comparisson}.

\begin{table}[t]
\begin{center}
\scalebox{0.9}{
\begin{tabular}{|c|c|c||c|c|c||c|c|c|}
\hline
\multicolumn{3}{|c||}{$\sigma_0=0.2,\>r=0.02$} &\multicolumn{3}{|c||}{$\sigma_0=0.5,\>r=0.05$} &\multicolumn{3}{|c|}{$\sigma_0=0.5,\>r=0.05$} \\
\multicolumn{3}{|c||}{$K=2,\>\ell=1.5,\>u=2.5$}  &  \multicolumn{3}{|c||}{$K=2,\>\ell=1.5,\>u=3$} & \multicolumn{3}{|c|}{$K=1.75,\>\ell=1,\>u=3$}   \\
\hline
GY & KI & MG & GY & KI & MG & GY & KI & MG \\
\hline
\hline
0.0411 & 0.041089 & 0.041082 & 0.0178 & 0.017856 & 0.017856 & 0.07615 & 0.076172 & 0.076165 \\
\hline
\end{tabular}
}
\end{center}
\vspace{2mm}
\caption{\footnotesize{The comparison of double barrier call option prices 
with strike
$K$
obtained 
in~\cite{GemanYor} and~\cite{KunitomoIkeda}
in the case of geometric Brownian motion.
The model is given by SDE~\eqref{eq:dif}
with the constant volatility function
$\sigma(x):=\sigma_0$
and drift
$\gamma=r-d$,
where 
the interest rate 
$r$
is given in the table and the dividend yield equals
$d=0$.
The asset price process 
$S$
starts at
$S_0=2$
and the maturity in all the cases is 
$T=1$
year.
The state-space of the approximating chain is
defined by the algorithm in Section~\ref{sec:MarkovChainConstr}
and the parameters
$N=200$,
$x_1= 0.2, x_N  = 10$,
$d_1^-= 100, 
d_1^+= 1,
d_2^- = 10, 
d_2^+=10, 
d_3^-=1, 
d_3^+=100$. 
The computation for the pricing of the barrier products takes 
about 
$0.03$
seconds
(using Matlab(R) 7.8 on Intel(R) Xeon(R) CPU E5430 @ 2.66GHz)
for each of the parameter choices 
in this table.
}}
\label{t:BS_Comparisson}
\end{table}
\begin{figure}[t]
\hspace{-0.6cm} 
\includegraphics[width=0.9\textwidth]{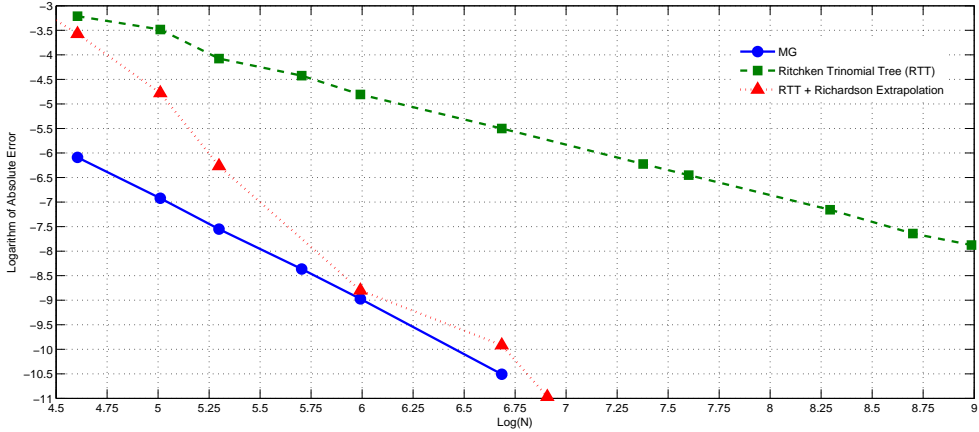}
\caption{\footnotesize{ 
This figure depicts the pricing errors for the double barrier
knock-out call 
option 
($S_0=95$,
$K=100$,
$\ell=90$,
$u=140$,
$T=1$,
$r=0.1$,
$\sigma_0=0.25$)
of the MG algorithm, 
the trinomial tree method described in Ritchken~\cite[Ex. 7]{Ritchken}
and its second order Richardson extrapolation.
The true value of this contract used to compute the pricing errors 
equals  1.4583798 
(this was 
obtained by the MG algorithm with $N=3000$ states). 
The figure is on a ``log-log'' scale where 
the vertical axes contains the logarithm of the absolute value of
the error and the horizontal axis depicts the logarithm of the number
of states 
$N$
in the grid. 
}}
\label{fig:Error_Bin}
\end{figure}
In Figure \ref{fig:Error_Bin} 
the errors and computation times of the Markov generator
algorithm (MG) and a trinomial tree are plotted against the number
of points, for the computation of the price of a double knock-out barrier call option. 
Note that the graph of the logarithm of the absolute 
pricing error for the MG algorithm 
given in
Figure~\ref{fig:Error_Bin}
is approximately linear 
in 
$\log(N)$
with slope
$-2$, which implies that
the error itself is approximately quadratic
in 
$1/N$.
The pricing algorithm, given in Ritchken~\cite[Ex. 7]{Ritchken},
is based on the trinomial tree and it appears to converge
at the rate
$1/N$.
Figure~\ref{fig:Error_Bin} shows 
that when the trinomial tree algorithm of Ritchken~\cite[Ex. 7]{Ritchken}
is combined with the second order Richardson extrapolation,
the convergence appears to be of higher order than in the MG algorithm.

\subsection{L\'evy and local L\'evy models}
\label{subsec:Levy}

\subsubsection{The CGMY/KoBoL process}
\label{subsec:CGMY}

In this section we assume that
the price 
$S$ 
is again an exponential L\'evy process
given by~\eqref{eq:Levy},
where the L\'{e}vy processes $L$ is 
a CGMY process 
\cite{CarrMadanGemanYor_CGMY} with
L\'evy density given by the formula
\begin{equation}
\label{eq:CGMY_LevyDensity}
k(y):=C\left( \I_{(-\infty,0)} \frac{\te{-G|y|}}{|y|^{Y+1}} + \I_{(0,\infty)} \frac{\te{-My}}{|y|^{Y+1}}\right),
\qquad\text{where}\quad \>M> 1\>,G\geq0,\>C>0,\>Y<2.
\end{equation}
The inequality  
$Y<2$
is induced by the integrability condition on the L\'evy measure
at zero
and the condition
$M>1$
implies
the exponential moment condition~\eqref{eq:Levy_Measure_Cond}.


Madan and Yor~\cite{MadanYor_CGMY_Sub} show that the CGMY process
$L$
has the same law as the time-changed Brownian motion 
$\{W_{Z_t} + \theta Z_t\}_{t\geq0}$ 
with 
$\theta=(G-M)/2$
and L\'evy 
subordinator 
$Z$
that is independent of 
$W$
and that has 
Laplace exponent 
$\psi_Z$
given by
\begin{eqnarray}
\label{eq:CGMY_Sub_LT}
& & 
\EE_0[\te{-u Z_t}] = \te{t\psi_Z(u)}   = 
\exp\left(tC\Gamma(-Y)[2r(u)^Y\cos(\eta(u) Y) - M^Y - G^Y]\right), \quad
u\geq-\frac{GM}{2},
\end{eqnarray}
where
$\Gamma$
denotes the Gamma function and 
the functions 
$r$,
$\eta$
are given by the formulae
$$r(u):= \sqrt{2u + GM}\quad \text{and} \quad
\eta(u):= \arctan\le(\frac{2\sqrt{2u - \theta^2}}{G+M}\ri).$$

We compare our numerical results with those obtained 
in Boyarchenko and Levendorskii~\cite{BoyLev_doublebarrier}  
using a Fourier method (see Table~\ref{table:BL}).

\begin{figure}[t]
\includegraphics[width=0.8\textwidth]{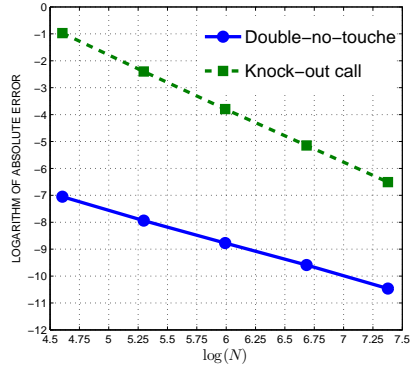}
\caption{\footnotesize{ 
The figure displays the logarithm of the absolute pricing errors for the MG algorithm applied to the CGMY model as a function of $\log(N)$, 
where $N$ is the number of states in the grid.
Under consideration  are the double no touch and double knock out call 
options with $K=S_0 =3500$, lower barrier $\ell=2800$ and upper barrier $u=4200$, driven by a CGMY process with parameters as in Table \ref{table:BL}. 
As true values we took the values $78.752$ and $0.9508$ obtained by running the algorithm with $N=6400$ points.
}}
\label{fig:CGMY}
\end{figure}

In Figure \ref{fig:CGMY} the error and computation times are plotted 
of the prices of double-no-touch  and double knock-out put options 
as a function of the grid size. As the true values we took the outcomes 
of the algorithm for $N=6400$. The figure appears to suggest that the error is approximately proportional to $N^{-1.2}$ and $N^{-2}$, 
respectively, where $N$ is the number of point in the grid. 

\begin{table}[t]
\centering
\scalebox{0.8}{
\begin{tabular}{ccc}
\begin{tabular}{|c|c|c||c|c|}
\hline
Spot & \multicolumn{2}{|c||}{Knock-out put} & \multicolumn{2}{|c|}{Double-no touch}\\
\hline
\% & BL & MG & BL & MG\\
\hline
82 & 302.28 & 301.07 & 0.5778 & 0.5757\\
85 & 370.38 & 370.38 & 0.8009  & 0.8004\\
88 & 341.35 & 341.78 & 0.8881 & 0.8880\\
91 & 279.86 & 280.41 & 0.9280 & 0.9280\\
94 & 207.71 & 208.30 & 0.9464 & 0.9465\\
97 & 136.63 & 137.24 &0.9527& 0.9529\\
100 & 78.19 & 78.74 & 0.9506 & 0.9507\\
\hline
\end{tabular} 
& & 
\begin{tabular}{|c|c|c||c|c|}
\hline
Spot & \multicolumn{2}{|c||}{Knock-out put} & \multicolumn{2}{|c|}{Double-no touch}\\
\hline
\% & BL & MG & BL & MG\\
\hline
101 & 64.07& 64.53 & 0.9481 & 0.9483\\
104 &36.96& 37.18 & 0.9351 & 0.9352\\
107 & 22.73 & 22.84 & 0.9112 & 0.9113\\
110 &14.60 & 14.65 & 0.8708 & 0.8709\\
113 & 9.61 & 9.64 &0.8020&  0.8019\\
116 &6.30& 6.32 &0.6771 & 0.6767\\
119 & 3.52 & 3.54 & 0.4049 & 0.4049\\
\hline
\end{tabular} 
\end{tabular}
}
\vspace{2mm}
\caption{\footnotesize{Barrier option prices under the CGMY model. The first
column contains the spot price as percentage of 3500. The CGMY
parameters are $C=1$, $G=9$, $M=8$, $Y=0.5$. The resulting
risk-neutral drift 
is $\mu=r-d-\psi_Z(-\theta-1/2)\approx -0.0423$. Option parameters
$K=3500$ (strike of the put), $\ell=2800$, $u=4200$, $r=0.03$, $d=0$,
$T=0.1$. The columns BL and MG report the results obtained by
Boyarchenko and Levendorskii~\cite{BoyLev_doublebarrier} and by 
the Markov generator algorithm (with 
$N=800$ points)
respectively.
It takes about 
$22$
seconds to run the MG algorithm for each starting spot price
}}
\label{table:BL}
\end{table}

\subsubsection{A local L\'evy model}
\label{subsec:Kou}
The following L\'evy driven SDE
specifies a Markov process 
$S$
with local volatility and 
double-exponential jumps,
\begin{eqnarray}
\label{eq:SDE_Kou}
\frac{\td S_t}{S_{t-}} & = & (r-d -\lambda\zeta (S_{t-}/S_0)^\beta) \td t+(S_{t-}/S_0)^\beta \td L_t,\q t>0,\qquad\text{where}\quad \\
L_t & := & \sigma_0W_t+ \sum_{i=1}^{N_t}\left(\te{K_i}-1\right),
\quad S_0, \sigma_0\in(0,\infty)\quad\text{and}\quad \beta\in\RR.
\label{eq:KouLevy}
\end{eqnarray}
The special case
of this model for
$\beta=0$
was introduced into the 
mathematical finance literature by Kou~\cite{Kou}.
The random variables 
$K_i$,
$i\in\NN$,
are independent of both the Brownian motion 
$W$
and the Poisson process
$N$
with intensity
$\lambda>0$
and are distributed according to the double exponential density
\begin{eqnarray}
\label{eq:DoubleExponentialDensity}
f_K(k) &  = &
p\eta_1 \te{-\eta_1 k}\I_{(0,\infty)}+
(1-p)\eta_2 \te{\eta_2 k}\I_{(-\infty,0)},\qquad\text{where}\\
& & \eta_1>1,\>\eta_2>0\quad\text{ and }\quad p\in[0,1].\nonumber
\end{eqnarray}
The parameter 
$\zeta$
is given by 
$$
\zeta:=\EE\left[\te{K_1}-1\right]=\frac{p\eta_1}{\eta_1-1}+\frac{(1-p)\eta_2}{\eta_2+1}-1.
$$
If $\beta<0$ the process $S$ has positive probability of hitting 
zero in finite time, in which case we take 0 to be absorbing.

It is clear that the model described by~\eqref{eq:SDE_Kou} and~\eqref{eq:KouLevy}
has a generator of the form given in~\eqref{eq:gengen}
with
$\sigma(x)=\sigma_0 (x/S_0)^\beta$
and
$\mu(x)=\lambda\zeta (x/S_0)^\beta$.
The jump measure 
$\nu(x,\td y)$
in representation~\eqref{eq:gengen}
is supported in
$(-1,\infty)$
and in our case 
by~\eqref{eq:DoubleExponentialDensity} 
takes the explicit form
\begin{eqnarray}
\label{eq:KouLevyMeasure}
\nu(x,\td y) & = & (x/S_0)^\beta\lambda\left[
p\eta_1 (y+1)^{-1-\eta_1}\I_{(0,\infty)}+
(1-p)\eta_2 (y+1)^{\eta_2-1}\I_{(-1,0)}\right]\td y.
\end{eqnarray}

Representation~\eqref{eq:KouLevyMeasure} of the 
jump measure 
$\nu(x,\td y)$
of the asset price process 
$S$
can now be used to construct 
the ``jump'' generator 
$\L_J$
defined in equations~\eqref{eq:JumpGen} and~\eqref{eq:JumpGenIsGen}.
Furthermore it is clear from~\eqref{eq:KouLevyMeasure}
that the instantaneous variance term caused by the jumps of the process
$S$
in~\eqref{eq:JumpGenEqu3}
is of the form
$$
\int_{-1}^\infty y^2\nu(x,\td y) = 
(x/S_0)^\beta2\lambda\left(\frac{p}{(\eta_1-1)(\eta_1-2)}+\frac{1-p}{(\eta_2+1)(\eta_2+2)}\right)\qquad
\text{if}\quad \eta_1>2.
$$
Note that in the present model the instantaneous variance is finite  (cf. condition~\eqref{eq:JumpSecondMomentCond}) if $\eta_1>2$
which is a condition that is typically satisfied in applications.
The numerical results of the MG
algorithm applied to an up-and-in call option
are contained in Table~\ref{t:Kou}, and are compared in the case 
$\beta=0$ with the corresponding results of Kou and Wang~\cite{Kou_Wang}
(see Table~3 in~\cite{Kou_Wang}).

\begin{table}[t]
\begin{center}
\scalebox{0.9}{
\begin{tabular}{|ccc||c|c|c||c|}
\hline
\multicolumn{3}{|c||}{Local L\'evy model} & MG: $N=400$  & MG: $N=800$  & MG: $N=1200$  
& KW \\
\hline
\hline
$\beta = 0$ & 
\multicolumn{2}{|c||}{ $\lambda=3$} & 10.0528  & 10.0530 & 10.0530 
& 10.05307\\
& \multicolumn{2}{|c||}{$\lambda=0.01$} & 9.2768 & 9.2771 & 9.2772 
& 9.27724 \\ 
\hline\hline
$\beta = -1$ &
\multicolumn{2}{|c||}{$\lambda=3$} & 9.7685 & 9.7688 & 9.7688 
& N/A\\
& \multicolumn{2}{|c||}{$\lambda=0.01$} & 8.9572 & 8.9575 & 8.9575 
& N/A \\ 
\hline\hline
$\beta = -3$ & 
\multicolumn{2}{|c||}{ $\lambda=3$} & 9.0185 & 9.0187 & 9.0188 
& N/A\\
&\multicolumn{2}{|c||}{$\lambda=0.01$} & 8.0855 & 8.0858 & 8.0858 
& N/A \\ 
\hline
\end{tabular}
}
\end{center}
\vspace{2mm}
\caption{\footnotesize{
Up-and-in call option prices under
the model 
$S$
defined in~\eqref{eq:SDE_Kou}
and~\eqref{eq:KouLevy}.
The parameters are given by
$S_0 = 100$,
$r = 5\%$,
$d=0$,
$\sigma_0 = 0.2$, 
$p = 0.3$,
$1/\eta_1 = 0.02$,
$1/\eta_2 = 0.04$,
$\beta\in\{0,-1,-3\}$
and
$\lambda\in\{0.01,3\}$.
The strike 
is
$K=100$
and
the upper barrier
$u=120$
while time to maturity
$T=1$.
The column KW 
denotes the results of 
Kou and Wang (see~\cite{Kou_Wang}, Table~3) 
while
MG denotes the algorithm based on the Markov generator.
The state-space of the approximating chain is
defined by the algorithm analogous to the one 
in Section~\ref{sec:MarkovChainConstr},
adapted in an obvious way to a single barrier contract.
Its size is
$N=n\cdot400$
for 
$n=1,2,3$.
The computation time is about half a second for
$n=1$,
five seconds for 
$n=2$
and seventeen seconds for 
$n=3$
on the same hardware as in 
Table~\ref{t:BS_Comparisson}.
We also ran the algorithm for 
$N=1600$
and obtained identical results (up to four decimals)
as the ones in the column
$N=1200$.
}}
\label{t:Kou}
\end{table}

\begin{figure}[t]
\includegraphics[width=0.8\textwidth]{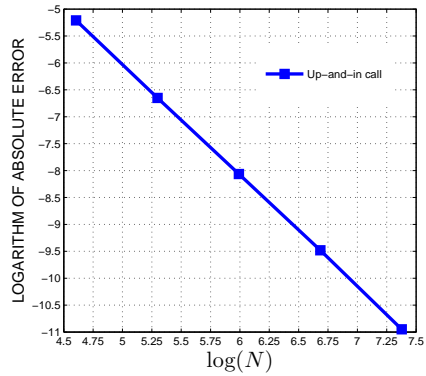}
\caption{\footnotesize{ 
The figure displays the logarithm of the absolute 
pricing errors for the MG algorithm applied to the 
model in~\eqref{eq:SDE_Kou} as a function of $\log(N)$,
where $N$ is the number of states in the grid.
The option under consideration is the
up and in call option specified in Table~\ref{t:Kou}.
We take the value
$9.768837$,
obtained by running the MG algorithm for $N=5000$, to be the correct prices in model~\eqref{eq:SDE_Kou} given 
by the parameter values $\lambda=3$ and $\beta= -1$.
In order to ensure that the truncation error is under control we computed the option value with the 
largest grid point equal to 300 and 600 (for $N=5000$) and obtained the same price as above in the first six decimal points.
%
%
%
}}
\label{fig:StateDepKou}
\end{figure}

Figure~\ref{fig:StateDepKou}
presents the pricing error for the up and in call option 
described in Table~\ref{t:Kou}.
The error of the pricing algorithm is plotted on a log-log-scale 
against the number of points in the state-space,
for the parameter values
$\lambda=3$
and
$\beta = -1$. 
Unlike in the case of double knock-out options where a truncation error
can be avoided, a truncation error will be present when pricing 
a single up-and-in option, as in the example above. To ensure 
that the truncation error is sufficiently small we ran 
the algorithm for an  increasing 
sequence of values of $x_N$, the largest point in the state-space 
$\mathbb G$ of the chain, and observed that the outcomes did not change 
up to the required accuracy. The slope of the line in Figure 
\ref{fig:StateDepKou} is equal to -2, which suggests that also in this case the algorithm is of order $h^2$ in the mesh size.

%% file: conclusion.tex
In this paper we presented an algorithm for pricing barrier options in 
one-dimensional 
Markovian models based on an approximation by continuous-time 
Markov chains. The generator of the approximating chain is constructed by 
the matching of instantaneous moments of the infinitesimal generator of 
the Markov process in question, on a suitable non-uniform grid.
The approximate barrier option prices are then obtained by calculating 
the corresponding first-passage distributions for the approximating 
Markov chain. 

To illustrate the flexibility of the method we 
implemented the algorithm for a number of models, including 
local volatility models with jumps and models 
with time-dependent jump-distributions 
(see \cite{MP_2010_LongVersion}). 
In the cases of the diffusion and jump-diffusion models 
where results had been obtained before in the 
literature, the algorithm produced outcomes that 
accurately matched those results, and we numerically 
investigated the order of decay of the error.  

We provided a mathematical proof of the 
convergence of the outcomes of the algorithm
to the true prices
and derived error estimates under additional regularity assumptions.
We derived a theoretical upper
bound for the error of the outcomes produced by the algorithm 
that is linear in the spatial mesh size and the truncation error.
We showed that an additional logarithmic factor may arise in this error bound
when the L\'evy density  
has a pole of order two at the origin.
In addition, this bound is also linear 
in the time mesh size 
if the model is time-inhomogeneous (see~\cite{MP_2010_LongVersion}). Numerical experiments  
suggest that for a number of models the error of the outcomes 
generated by the algorithm actually decays quadratically in the spatial mesh-size.
It would be of interest to establish error bounds under weaker 
regularity assumptions, and obtain sharp rates of convergence 
for the specific models, which is a topic left for future research.

Although in principle the method also applies to higher-dimensional 
Markov processes, the size of 
the generator matrix would make straightforward application 
of the algorithm computationally infeasible. 
The investigation of efficient extensions of the approach 
to Markov processes of moderate dimension is another topic 
left for future research.

%% file: proofs.tex
\subsection{Explicit construction of the generator matrices $\Lambda^{(n)}$}\label{a:MC}

Recall that  the generator matrix 
$\Lambda^{(n)}$
is of the form
$\Lambda^{(n)}_D+\Lambda^{(n)}_J$.
The action of $\Lambda^{(n)}_D$,
which discretizes the operator~\eqref{eq:L_Diff}, on a function 
$f:\mathbb G^{(n)}\to\mathbb R$ is described 
 by  
\begin{eqnarray}\label{LDx}
\Lambda^{(n)}_Df(x_i) = 
\begin{cases}
\frac{\sigma^2_{i}}{2}\Delta f(x_i) + (\gamma_{i})_+
\nabla^+ f(x_i) + (\gamma_{i})_- \nabla^- f(x_i) & i\in\{2, \ldots, n-1\},\\
0 & i\in\{1,n\},
\end{cases}
\end{eqnarray}
where $y_\pm=\max\{\pm y,0\}$ and $\mbb G^{(n)}$ is 
defined in \eqref{eq:GT}. 
We write $x_i$ instead of $x^{(n)}_i$ and will do so in all that 
follows. 
We denote by $\nabla^\pm$ and $\Delta$ 
the first and second order difference operators given by
\begin{eqnarray}
\label{nabla-}
\nabla^-f(x_i) &=& \frac{1}{x_i - x_{i-1}}[f(x_i) - f(x_{i-1})],\\
\label{nabla+}
\nabla^+f(x_i) &=& \frac{1}{x_{i+1} - x_{i}}[f(x_{i+1}) - f(x_{i})],\\
\Delta f(x_i) &=& \frac{2}{x_{i+1}-x_{i-1}} [\nabla^+ f - \nabla^- f](x_i),
\label{delta}
\end{eqnarray}
with $\sigma^2_{i}$ and $\gamma_{i}$ defined by
$$
\sigma_{i}^2 := \sigma^2(x_i)x_i^2,\qquad\q \gamma_{i} := \gamma x_i,
$$
where $\sigma^2(x)$ and $\gamma$ are as in \eqref{eq:gengengen}.
Note that the discretization in~\eqref{LDx} 
defines a birth-death process also in the case that 
$\sigma_{i}=0$.

The jump part of the generator, given in~\eqref{eq:L_Jump},
can be rewritten 
as follows
\begin{eqnarray}
\label{eq:RepOfJumpGen}
\lefteqn
{\mathcal L_{J}f(x)= \int_{1}^{\infty}[f(x(1+y)) - f(x) -f'(x)xy]g(x,y) \td y}\\
&+ & 
\begin{cases}
\int_{-1}^{1} \frac{1}{|y|}[f(x(1+y)) - f(x)] |y|g(x,y) \td y
-\> f'(x)x\int_{-1}^{1} yg(x,y) \td y,& \quad\mc I(x)  < \infty,\\
\int_{-1}^{1}\frac{1}{y^2}[f(x(1+y)) - f(x) - f'(x)xy](y^2g(x,y)) \td y,& \quad\mc I(x)  = \infty.\\
\end{cases}
\nn
\end{eqnarray}
Note that
$\mc I(x)$,
defined in~\eqref{Itx},
is either finite 
for all
$x$
or infinite 
for all
$x$
because we are assuming that the process
$S$
we are approximating falls into  one of the categories  O, I, II.
The representation~\eqref{eq:RepOfJumpGen} has a natural discretization,
which we will now describe, because the second integral
in~\eqref{eq:RepOfJumpGen}
is an integral of a continuous  function against 
a finite measure.

For different levels of activity of the purely discontinuous part of $S$,
quantified by whether or not $\mathcal I_{i}= \mc I(x_i)$ 
defined in \eqref{Itx} is finite, 
the matrix $\Lambda_J^{(n)}$ is defined in the following way.
Let
$$A^0_i(h) = (-1 + x_{\unl b_i}/x_i, -1 + x_{\ovl b_i}/x_i)$$
where the levels $\unl b_i$ and $\ovl b_i$ are given by 
\begin{equation}
\ovl b_i = \min\{j>i: x_j\ge x_i + C_+(h) h\}, \q\q
\unl b_i = \max\{j<i: x_j\leq x_i - C_-(h) h\},
\end{equation}
with $h=h(n)$ the spatial mesh size and
\begin{equation}\label{Cpm}
C_\pm(h) = d_\pm \cdot c(\alpha_\pm,h),
\q\q
\q\text{for some constants}\ d_\pm > \frac{\ovl\kappa_\pm}{\unl\kappa_\pm},
\end{equation}
where $\alpha_\pm$, $\ovl \kappa_\pm$ and $\unl\kappa_\pm$ 
are the constants given in 
\eqref{stable-up} and \eqref{stable-down}, and 
\begin{equation}
c(\a,h) = 
\begin{cases}
\displaystyle\frac{2-\alpha}{\alpha-1}, & \a\in(1,2),\\
-\log(h), & \alpha = 1.
\end{cases}
\end{equation} 
We can then define
$$
c^{(p)}_{ik} = 
\begin{cases}
\int_{\mathbb R} |y|^p g(x_i,y)\I_{\le[\frac{x_{k-1}}{x_i}-1,%
 \frac{x_{k}}{x_i}-1\ri]}(y)\td y
,& k>i,\\
\int_{\mathbb R} |y|^pg(x_i,y) \I_{A^0_i(h)}(y)\td y
,& k=i,\\
\int_{\mathbb R} |y|^p g(x_i,y)\I_{\le[\frac{x_{k}}{x_i}-1,%
 \frac{x_{k+1}}{x_i}-1\ri]}(y)\td y,& k<i.
\end{cases}
$$
Finally 
we  define the discretized version of the generator
as
\begin{eqnarray}\label{eq:LJ}
\Lambda_J^{(n)} f(x_i) &=& 
\sum_{k:x_k>2x_i} \left[c^{(0)}_{ik}(f(x_k)-f(x_i)) - 
x_i\nabla^- f(x_i)c^{(1)}_{ik}\right] +  \overline{\Lambda}_J^{(n)} f(x_i),\>
i\notin\{1,n\},\\
\nn \Lambda_J^{(n)} f(x_i) &=& 0, \q i\in\{1,n\}.
\end{eqnarray}
If 
$\mathcal I_{i}$
is finite 
$\overline{\Lambda}_J^{(n)} f(x_i)$ 
is given by
\begin{eqnarray}
\label{eq:Jump_Gen_Finite_Var}
\overline{\Lambda}_J^{(n)} f(x_i) 
&:=&  
\sum_{k\neq i:x_k\leq 2x_i} \frac{x_i}{|x_k-x_i|} \left[f(x_k)-f(x_i)\right]c^{(1)}_{ik}  - 
x_i\alpha_{i} 
\WT{\nabla}f(x_i)
\end{eqnarray}
where
\begin{eqnarray}
\label{eq:nabla_tilda}
\WT{\nabla}f(x_i) &:=&
\begin{cases}
\nabla^- f(x_i)& \alpha_{i} \geq0\\
\nabla^+ f(x_i)& \alpha_{i}  <0
\end{cases}
\quad\text{where}\qquad
\alpha_{i}:= \int_{-1}^1y g(x_i,y)\td y. 
\end{eqnarray}
If 
$\mathcal I_{i}$
is equal to infinity
we have
\begin{eqnarray}
\overline{\Lambda}_J^{(n)} f(x_i) \nonumber
&:=& 
\frac{1}{2}c_{ii}^{(2)}x_i^2\Delta f(x_i)\\
& & +
\sum_{k:(x_k-x_i)/x_i\in (-1,1)\backslash A_i^0(h)} \left(\frac{x_i}{x_k-x_i}\right)^2\left[f(x_k)-f(x_i) - (x_k-x_i)\nabla^0 f(x_i)\right]
c^{(2)}_{ik}.
\label{eq:Jump_Gen_Infinite_Var}
\end{eqnarray}
Here $\nabla^0f$ is the (central) difference
operator given by 
\begin{equation}\label{nabla0}
\nabla^0 f(x_i) = \frac{x_i - x_{i-1}}{x_{i+1} - x_{i-1}}
\nabla^+ f(x_i) 
+ \frac{x_{i+1} - x_i}{x_{i+1} - x_{i-1}}\nabla^- f(x_i).
\end{equation}

Clearly, the first part in the expression in \eqref{eq:LJ} 
defines the generator matrix of a Markov chain
taking values in $\mbb G^{(n)}$. 
Similarly the expression in~\eqref{eq:Jump_Gen_Finite_Var}
yields a generator of a Markov chain.
In~\eqref{eq:Jump_Gen_Infinite_Var} 
the levels 
$\ovl b_i$ 
and 
$\unl b_i$ 
are chosen so as to ensure 
that 
$\Lambda_J^{(n)}$ defined in \eqref{eq:LJ} 
is in fact a generator.

To verify that \eqref{eq:Jump_Gen_Infinite_Var} defines a Markov jump process we need to 
show that in the matrix $\Lambda_J^{(n)}$ defined in  
\eqref{eq:LJ} all off-diagonal elements are positive. 
It follows from the definitions of 
$\Delta f$
and
$\nabla^0 f$
that this is the 
case if the following condition holds for all 
$x_i\in\WH{\GG}^{(n)}$:
\begin{eqnarray}\label{s.cond}
x_i c^{(2)}_{ii}&\ge &
 \max\{(x_i-x_{i-1})\nu_{i},
(x_{i}-x_{i+1})\nu_{i}\}, \q\text{where}\\ 
\nu_{i}&:=&\sum_{k:(x_k-x_i)/x_i\in (-1,1)\backslash A_i^0(h)}\frac{x_i}{x_k-x_i}c_{ik}^{(2)}.
\end{eqnarray}
For 
$h$
sufficiently small
it is easily checked from the definition of $\nu_{i}$ that
$$
\nu_{i} \leq \int_{(-1,1)\backslash A_i^0(h)} |y| g(x_i,y)\td y,
$$
which implies that \eqref{s.cond} is satisfied if,
for any  $x\in[\ell,u]$,
$$
x \int_{A^0_i(h)}y^2 g(x,y)\td y
\ge h 
\int_{(-1,1)\backslash A_i^0(h)}|y|g(x,y)\td y.
$$
This condition is satisfied if, for  
any $x\in [\ell,u]$, the following 
two conditions hold 
for $h$ sufficiently small
\begin{eqnarray}\label{condp}
x\int_{\le(0,\frac{C_+(h)h}{x}\ri)} y^2g(x,y)\td y &\ge& h 
\int_{\le(\frac{C_+(h)h}{x},1\ri)}yg(x,y)\td y, \\
x\int_{\le(-\frac{C_-(h)h}{x},0\ri)} y^2g(x,y)\td y &\ge& h 
\int_{\le(-1,-\frac{C_-(h)h}{x}\ri)}|y|g(x,y)\td y.
\label{condm}
\end{eqnarray}
In view of \eqref{stable-up} 
it follows that \eqref{condp} holds if
 the following inequality is satisfied
\begin{eqnarray*}
\lefteqn{\unl\kappa_+\int_{(0, C_+(h)h/x)}y^{1-\a_+}\td y \ge h \ovl\kappa_+
\int_{(C_+(h)h/x, 1)} y^{-\alpha_+}\td y} \\
&& \Leftrightarrow  
\begin{cases}\displaystyle
\frac{\unl\kappa_+ x}{2-\alpha_+}
\le(\frac{C_+(h) h}{x}\ri)^{2-\alpha_+} \ge \frac{\ovl\kappa_+ h}{\alpha_+-1}
\le(\le(\frac{C_+(h) h}{x}\ri)^{1-\alpha_+} - 1\ri) & \text{in case II},\\
\unl\kappa_+x\le(\frac{C_+(h) h}{x}\ri)
\ge -\ovl\kappa_+ h\le(\log(C_+(h)) + \log(h) - \log(x)  \ri) 
& \text{in case I.}
\end{cases}
\end{eqnarray*}
The latter holds for all $h$ sufficiently small if 
$$
\begin{cases}
\displaystyle C_+(h)\cdot\frac{\unl\kappa_+}{2-\alpha_+} 
\ge \frac{\ovl\kappa_+}{\a_+ - 1},& \text{in case II},\\
\liminf_{h\to\infty} C_+(h)/|\log h| 
> \ovl\kappa_+/\unl\kappa_+,
& \text{in case I}.
\end{cases}
$$
Clearly, $C_+(h)$, defined in \eqref{Cpm},  satisfies this condition. 
The fact that Condition \eqref{condm} is satisfied 
follows by a similar line of reasoning.

\subsection{Proof of error estimates}\label{a:error}

The following lemma is an important auxiliary result for the development
of error-estimates. It formalises the intuition that 
two semi-groups should be ``close'' if the corresponding 
infinitesimal generators are ``close''.

Suppose that $S$ is a Feller process on the state-space $\E$ 
with associated semigroup $P$ 
and corresponding infinitesimal generator $\mathcal L$, and 
let $(X^{(n)})_{n\in\mathbb N}$ 
be a sequence of Markov chains with state-spaces 
$\GG^{(n)}$, semigroups $P^{(n)}$ and corresponding 
generator matrices $\Lambda^{(n)}$. 
For any $f\in\mathcal D(\mathcal L)$, the domain of $\mathcal L$,
consider the following error measure
\begin{equation}\label{errornorm}
\epsilon_n(f) = \left\|\Lambda^{(n)}f - 
\mathcal Lf\right\|_n, \qquad s\ge 0,
\end{equation}
where we write $\Lambda^{(n)}f = \Lambda^{(n)} f_n$ and $P^{(n)} f = P^{(n)}f_n$
with $f_n = f|_{\mathbb G^{(n)}}$ and 
\begin{equation}\label{norm}
\|g\|_n = \sup_{x\in\mathcal\GG^{(n)}}|g(x)|
\end{equation}
for any $g:\mathbb E\to\mathbb R$.
Then the following estimate holds true:

\begin{Lemma}
\label{lem:gen-estimate0}
Let $f\in\mathcal D(\mathcal L)$ and $T>0$, and suppose that there 
exists a function $h:\mathbb N\to\mathbb R_+$ and a function 
$c_f:\mathbb R_+\to\mathbb R_+$ such that for 
all $n\in\mathbb N$ and $s\ge 0$
\begin{equation*}
\epsilon_n(P_sf) \leq c_f(s) \cdot h(n).
\end{equation*}
Then it holds that
\begin{equation*}
\sup_{t\in[0,T]}\le\|P_t f - P^{(n)}_t f\ri\|_n
\leq \int_0^Tc_f(s)\td s \cdot h(n).
\end{equation*}
\end{Lemma}

\proof The proof closely follows that of Lemma 6.2 in Ethier and Kurtz \cite{EthierKurtz}.
Note first that 
\begin{eqnarray}\nn
\frac{\partial}{\partial s}\le(P^{(n)}_{t-s}P_s f\ri) &=& 
\Lambda^{(n)} P^{(n)}_{t-s}P_s f - P^{(n)}_{t-s}\mathcal L P_s f\\
&=& P^{(n)}_{t-s}\le(\Lambda^{(n)} P_s f - \mathcal L P_s f\ri),
\label{eq:pn}
\end{eqnarray}
as $P^{(n)}_{t-s}$ and $\Lambda^{(n)}$ commute.
Since the function in \eqref{eq:pn} is continuous in $s$, 
the fundamental theorem of 
calculus and the fact that $P_0^{(n)}f=f$ imply that
$$
P_t f - P^{(n)}_t f = \int_0^t P^{(n)}_{t-s}
\le(\Lambda^{(n)} P_s f - \mathcal L P_s f\ri)\td s.
$$
Thus, the triangle inequality yields that
\begin{eqnarray*}
\left\|P_t f - P^{(n)}_t f\right\|_n &\leq& \int_0^t \le\| P^{(n)}_{t-s}\le(\Lambda^{(n)} P_s f - \mathcal L P_s f\ri)\ri\|_n\td s\\
&\leq & \int_0^t \le\|\Lambda^{(n)} P_s f - \mathcal L P_s f\ri\|_n\td s\\
&\leq & \int_0^T c_f(s)\td s \cdot h(n)
\end{eqnarray*}
for any $t\in[0,T]$, as $c_f(s)$ is non-negative. \exit

Before we proceed we introduce some further notation.
The integral part of the infinitesimal generator, and its approximation
will be denoted by
\begin{eqnarray}
\label{eq:If}
I f(x) &=& \int_{(1,\infty)}[f(x(1+y))-f(x)- xyf'(x)]g(x,y)\td y,\\
I_n f(x_i) &=& \int_{(1,x_n/x_i - 1)}
[f_n(x_i(1+y)) - f_n(x_i)]g(x_i,y)\td y - 
\mu_{i}x_i\nabla^- f(x_i), 
\end{eqnarray} 
where $\mu_i=\int_{(1,\infty)}yg(x_i,y)\td y$,
$'$ denotes the derivative with respect to $x$ and
$$f_n(x) = \sum_{x_k\in(2x_i,\infty)}
f(x_{k+1})\I_{\le[\frac{x_{k}}{x_i}-1,\frac{x_{k+1}}{x_i}-1\ri]}(x).$$
Furthermore, we define  
\begin{eqnarray}
\label{eq:Istar}
I^{(k)} f(x) &=& \int_{(-1,1]}
F^{(k)}(x,y) g(x,y)|y|^k\td y, \\
I^{(k)}_n f(x_i) &=& \int_{(-1,1]}F^{(k)}_{n,i}(y) g(x_i,y)|y|^k\td y,
\label{eq:Instar}
\end{eqnarray}
where, for $k=1,2$, 
\begin{eqnarray*}
F^{(k)}(x,y) &=& |y|^{-k}[f(x(1+y))-f(x) - f'(x)xy],\\
F^{(1)}_{n,i}(y) &=& -x_i \WT{\nabla} f(x_i)\text{sgn}(y)+\sum_{x_j\in [0,2x_i]\backslash\{x_i\}} F_{n,ij}^{(1)}
\I_{\le[\frac{x_{j}}{x_i}-1,\frac{x_{j+1}}{x_i}-1\ri]}(y),\\
F^{(2)}_{n,i}(y) &=& 
\sum_{(x_j/x_i-1)\in (-1,1)\backslash A_i^0(h)} \le[ F_{n,ij}^{(2)}
-\frac{x_i^2}{x_j-x_i}\nabla^0 f(x_i)\ri] 
\I_{\le[\frac{x_{j}}{x_i}-1,\frac{x_{j+1}}{x_i}-1\ri]}(y)\\
& &+
\frac{x_i^2}{2}\Delta f(x_i)\I_{A_i^0(h)}(y), 
\end{eqnarray*}
where
$\WT{\nabla}$
is defined in~\eqref{eq:nabla_tilda},
with
$$
F_{n,ij}^{(1)} = \frac{x_i}{|x_j-x_i|}
\begin{cases}
[f(x_{j+1}) - f(x_i)], 
& j>i,\\
[f(x_{j})-f(x_i)],
& j<i,
\end{cases}
$$
and 
\begin{eqnarray*}
F^{(2)}_{n,ij} &=& 
\le(\frac{x_i}{x_j-x_i}\ri)^2
\begin{cases}
[f(x_{j}) - f(x_i)], 
&\mbox{$\frac{x_j}{x_i}$}-1\in (-1,0)\backslash A_i^0(h),\\
[f(x_{j+1})-f(x_i)],
&\mbox{$\frac{x_j}{x_i}$}-1 \in(0,1)\backslash A_i^0(h).
\end{cases}
\end{eqnarray*}
%
%
%

\noindent{\it Proof of Theorem \ref{prop:error}.} 
In view of Assumption \ref{as:f}, it follows that the stopped 
and discounted
process  $\WT S^A = \{\te{-r(t\wedge\tau_A)}S_{t\wedge\tau_A}\}_{t\ge0}$  is a Feller process.
%
The infinitesimal generator  of the 
stopped semi-group $\WT P^A$ corresponding to $\WT S^A$
is denoted by $\WT{\mathcal L}^{A}$.
The process $\WT S^A$ is approximated by the Markov chains 
with infinitesimal generators $\WT\Lambda^{(n)}_r$, 
where $\Lambda^{(n)}$ is described above and $\WT{\Lambda}^{(n)}_r$ 
is obtained from $\Lambda^{(n)}$ 
by setting the rows of $\Lambda^{(n)}$ 
corresponding to $A$ equal to zero 
and subtracting $r$ on the diagonal elements of the rows corresponding 
to $A^c$, as in \eqref{eq:Def_tilA}.

For any $f$ satisfying Assumption \ref{as:f}, 
$\WT P_tf$ has compact support as 
$\WT P_tf(x)=f(x)$ for all $x\in A$ (recall we take $u<\infty$). 
Furthermore also by Assumptions \ref{as:bounded} and \ref{as:f} and the form of the Feller process $S$ under consideration $\WT P_tf$
is contained in the domain $\mc D$ of the infinitesimal 
generator $\mc L$. Since $$\te{-r(s\wedge\tau_A)}\WT P_tf(S_{s\wedge\tau_A}) = E[\te{-r((s+t)\wedge\tau_A)} f(S_{(s+t)\wedge\tau_A})|\mc F_s]$$ 
is a martingale, Dynkin's 
lemma implies that  $$(\mc L - r)(\WT P_tf)(x) =  0, \qquad x\in A^c,$$ 
so that in particular condition \eqref{eq:boundary} holds true 
with $f$ replaced by $\WT P_tf$.
From Lemma \ref{lem:KeyLemma} and Assumption \ref{as:f} we find that
$\WT{\mathcal L}^A(\WT P_tf)$ is given by \eqref{eq:LA}. 
Recall that, in the case under consideration, 
$\mc L f$ is of the form \eqref{eq:gengen}.

The form of the approximation and the triangle inequality imply 
that, for $H=\WT P_s f$, $s\in[0,T]$, 
the distance $\e'_n(H)$ between $\mc L$ and $\Lambda^{(n)}$
can be estimated as
\begin{eqnarray}\nn
\e'_{n}(H):=
\le\|\WT\Lambda^{(n)}_rH - (\WT{\mathcal L}^{A}-r)H\ri\|_n  &\leq& \mbox{$\frac{1}{2}$}\|\Sigma^2\|
\|\Delta_n H - H''\|
+ \|\Gamma\|\|\nabla_n H - H'\|
+ \|I H - I_n H\|\\
&+& \nonumber
\begin{cases}
\|I^{(1)}H - I^{(1)}_n H\| & \text{ if}\q\mc I_{i,j}<\infty,\\
\|I^{(2)}H - I^{(2)}_n H\| & \text{ if}\q\mc I_{i,j}=\infty,
\end{cases}
\label{eq:genest}
\end{eqnarray}
where 
$\mc I_{i,j}$
is defined in~\eqref{Itx}.
The norm $\|\cdot\|_n$ was defined in \eqref{norm} and
we wrote $\|\cdot\|=\sup_{x\in\WH{\mathbb G}^{(n)}}|\cdot|$ 
for the supremum over all the points in the grid that lie between
the barriers. The functions 
$\Sigma$
and 
$\Gamma$
are given by
$\Sigma^2(x)=\sigma^2(x)x^2$
and 
$\Gamma(x)=\gamma x$
where
$\gamma=r-d$.
The operator
$\Delta_n$ is equal to $\Delta$ defined in \eqref{delta} and 
$\nabla_n$ is given by $\nabla^+$ 
or $\nabla^-$ in \eqref{nabla+} and \eqref{nabla-} according to whether the constant
$\gamma$ is positive or negative, 
as in \eqref{LDx}. 
We denote by $H'$ the derivative of $H$ with respect to $x$.

Writing 
$$
 \|\cdot\|_{\infty}= \sup_{x\in\E}|\cdot|,
$$
%
%
%
%
Assumptions  \ref{as:bounded} and \ref{as:f}
and second and third order Taylor expansions 
yield that 
%
%
\begin{eqnarray*}
\|\nabla_n H - H'\| &\leq& \frac{h(n)}{2}\|H''\|_\infty\\
\|\Delta_n H - H''\| &\leq& \frac{2h(n)}{3}\|H'''\|_\infty.
\end{eqnarray*}
Furthermore, the triangle inequality 
and Assumptions  \ref{as:bounded} and \ref{as:f}
yield that,  for 
$h(n)$
sufficiently small, we have
\begin{eqnarray*}
\|IH - I_n H\| 
&\leq& \max_{x_i\in\WH\GG^{(n)}}\le\{ \sum_{x_{k}\in(2x_i,x_{n}]}
\int_{\le(\frac{x_k}{x_i}-1,\frac{x_{k+1}}{x_i} - 1\ri)}
\le|H(x_i(1+y)) - H(x_{k+1})\ri| g(x_i,y)\td y\ri.\\
&+&\le. \int_{\le(1,1+\frac{h(n)}{x_i}\ri]\cup\le[\frac{x_{n}}{x_i }-1, \infty\ri)}H(x_n(1+y)) g(x_i,y)\td y
+ |\mu_i|x_i \le|H'(x_i)-\nabla^-H(x_i)\ri|\ri\}
\\
&\leq& h(n) \cdot \|H'\|_\infty \cdot 
\max_{x_i\in\WH\GG^{(n)}}\int_{(1,\infty)}g(x_i,y)\td y \\
&+&
\|H\|_{\infty} \max_{x_i\in\WH\GG^{(n)}} \int_{\le(1,1+\frac{h(n)}{x_i}\ri]\cup\le[\frac{x_{n}}{x_i }-1, \infty\ri)}g(x_i,y)\td y 
+ \frac{h(n)}{2}\|H''\|_\infty \max_{x_i\in\WH\GG^{(n)}} |\mu_i|x_i 
\\
&\leq& C' \|H'\|_\infty \cdot h(n) + \|H\|_{\infty} \cdot \le(k(n)+\frac{h(n)}{\ell}\cdot\text{const}\ri)
+ \frac{h(n)}{2}\|H''\|_\infty C' u, 
\end{eqnarray*}
where $C'$ is given in Assumption \ref{as:bounded}, 
$\mu_i=\int_{(1,\infty)}yg(x_i,y)\td y$,
and $h(n)$ and $k(n)$ 
are the spatial mesh size and tail-mass defined 
in \eqref{eq:hd} and \eqref{eq:k},
and \texttt{const} is a bound for 
$g$
on
$[\ell,u]\times [1,2]$.
In case O the triangle inequality yields that,
for 
$n$
sufficiently large, 
$$
\|I^{(1)} H - I^{(1)}_n H\| \leq \WT C_0
\le(\|H''\|_{\infty}+\|H'\|_{\infty}\ri)
h(n)
$$
for some positive constant
$\WT C_0$.
Indeed,
\begin{eqnarray}
\lefteqn{\|I^{(1)} H - I^{(1)}_n H\|}\\ 
&\leq& 
\nn
\max_{x_i\in\WH{\GG}^{(n)}}\le\{
\sum_{x_k\in[0,2x_i]\backslash\{x_i\}}\int_{\le(\frac{x_k}{x_i}-1, 
\frac{x_{k+1}}{x_i} - 1\ri)} \le|F^{(1)}(x_i,y) - F^{(1)}(x_i,
\mbox{$\frac{x_k}{x_i}$} - 1)\ri| g(x_i,y)y\td y\ri. \\
\nn
&+& 
x_i \le|\int_{(-1,1]}y g(x_i,y)\td y\ri|\cdot \le|H'(x_i)-\WT{\nabla}H(x_i)\ri|\\
&+& 
\nn
\le. \int_{\le(0,\frac{x_{i+1}}{x_i} - 1\ri)} \le|F^{(1)}(x_i,y)
-x_i\WT{\nabla}H(x_i)\ri|g(x_i,y)y\td y\ri\}\\
& =:& \max_{x_i\in\WH{\GG}^{(n)}}\{J_1(x_i) + J_2(x_i) + J_3(x_i)\}.
\nn
 \label{eq:I1}
 \end{eqnarray}
In view of Assumptions  \ref{as:bounded} and \ref{as:f}
the three terms in \eqref{eq:I1} can be estimated as follows:
\begin{eqnarray*}
J_1(x_i) &\leq& \sum_{x_k\in(0,2x_i], k\neq i}
\int_{\le(\frac{x_k}{x_i}-1, 
\frac{x_{k+1}}{x_i} - 1\ri)}\le[ \sup_{y\in\le(\frac{x_k}{x_i}-1, 
\frac{x_{k+1}}{x_i} - 1\ri)}\le|\frac{\partial F}{\partial y}(x_i,y)\ri| \frac{x_{k+1} - x_k}{x_i} \ri] g(x_i,y)y\td y\\
&\leq&  \sum_{x_k\in(0,2x_i], k\neq i}\int_{\le(\frac{x_k}{x_i}-1, 
\frac{x_{k+1}}{x_i} - 1\ri)} g(x_i,y)y\td y\cdot  
\|H''\|_\infty \cdot x_i \cdot h(n) \\
&\leq& \frac{3u}{2}\cdot C'\cdot \|H''\|_\infty \cdot h(n),
\end{eqnarray*}
using that $|\frac{\partial F^{(1)}}{\partial y}(x,y)|\leq \frac{3x^2}{2} \|H''\|_\infty$, by a Taylor expansion.
Furthermore,
for 
$h(n)$
sufficiently small,
we have
\begin{eqnarray*}
J_2(x_i) 
&\leq& u\cdot \frac{h(n)}{2}\cdot \|H''\|_\infty 
\int_{\le(-1,1\ri)} g(x_i,y)|y|\td y\\
&\leq& \frac{u}{2}\cdot\|H''\|_\infty C_0  \cdot h(n),\\
J_3(x_i) &\leq& \frac{x_{i+1} - x_i}{x_i} \le(\frac{x_i^2}{2} \|H''\|_{\infty}
+x_i \|H'\|_{\infty}\ri)
\int_{(0,\frac{x_{i+1}}{x_i} - 1)}y g(x_i,y)\td y\\ 
&\leq&
h(n)\cdot \le(\frac{u}{2}\|H''\|_{\infty}+\|H'\|_{\infty}\ri)\cdot C_0,
\end{eqnarray*}
where we used that for $y\in(x_i,x_{i+1})$,
$|F^{(1)}(x_i,y)|\leq \frac{x_i^2}{2}\|H''\|_{\infty}$, by a Taylor expansion.

In  cases I and II, 
\begin{eqnarray}
 \nn
\lefteqn{\|I^{(2)} H - I^{(2)}_n H\|}\\
\nn&\leq& 
\max_{x_i\in\WH{\GG}^{(n)}}\le\{
\sum_{x_k\in (-1,1)\backslash A_i^0(h)}
\int_{\le(\frac{x_k}{x_i}-1, \frac{x_{k+1}}{x_i} - 1\ri)} \le|F^{(2)}(x_i,y) - F^{(2)}(x_i,
\mbox{$\frac{x_k}{x_i}$} - 1)\ri| g(x_i,y)y^2\td y\ri. \\
\nn
&+& 
\sum_{x_k\in (-1,1)\backslash A_i^0(h)}
\frac{x_i^2}{|x_k-x_i|} \le|H'(x_i)-\nabla^0H(x_i)\ri|\cdot\int_{\le(\frac{x_k}{x_i}-1, \frac{x_{k+1}}{x_i} - 1\ri)}y^2 g(x_i,y)\td y \\
\nn
&+& \le. \int_{A_i^0(h)} \le|F^{(2)}(x_i,y) - \mbox{$\frac{x_i^2}{2}$} \Delta H(x_i)\ri|
g(x_i,y)y^2\td y\ri\}\\
&=:& \max_{x_i\in\WH{\GG}^{(n)}}\{K_1(x_i) + K_2(x_i) + K_3(x_i)\}.
 \label{eq:I2}
\end{eqnarray}
In view of Assumptions  \ref{as:bounded} and \ref{as:f}
the three terms in \eqref{eq:I2} can be estimated as follows:
\begin{eqnarray*}
\lefteqn{K_1(x_i)}\\ 
&\leq& \sum_{(x_k/x_i-1)\in (-1,1)\backslash A_i^0(h)} \frac{x_{k+1}-x_k}{x_i} 
\cdot \max_{y\in \le(\frac{x_k}{x_i} - 1, \frac{x_{k+1}}{x_i} - 1\ri)}\le|\frac{\partial F^{(2)}}{\partial y}(x_i,y)\ri|
\cdot \int_{\le(\frac{x_k}{x_i} - 1, \frac{x_{k+1}}{x_i} - 1\ri)}
g(x_i,y)y^2\td y\\
&\leq& \frac{5}{6\ell} \cdot C'' \cdot \|H'''\|_\infty  \cdot h(n),
\end{eqnarray*}
where 
$C''$
is a constant that satisfies
$C'' \leq 3 C'$
for 
$h$
sufficiently small
and the bound
$$
\frac{\partial F^{(2)}}{\partial y}(x,y) = 
\frac{\partial}{\partial y}\frac{H(x(1+y)) - H(x) - H'(x)xy}{y^2}
\leq 
\frac{5 x^3}{6}\cdot
\max_{z\in[1\wedge(1+y),1\vee(1+y)]}|H'''(xz)|  
$$
holds.
Furthermore
we find
\begin{eqnarray*}
K_2(x_i) 
&\leq & x_i^2 \cdot \|H'''\|_\infty
\cdot \sum_{(x_k/x_i-1)\in (-1,1)\backslash A_i^0(h)} \le|\mbox{$\frac{(x_{i+1}-x_i)(x_i-x_{i-1})}{x_k-x_i}$}\ri|
\int_{\le(\frac{x_k}{x_i}-1,\frac{x_{k+1}}{x_i} - 1\ri)}g(x_i,y)y^2\td y\\
&\leq&  u^2\|H'''\|_\infty \cdot C'' \cdot h(n).
\end{eqnarray*}
Note that the equality
$$
F^{(2)}(x_i,y) - \mbox{$\frac{x_i^2}{2}$} \Delta H(x_i) =
\frac{x_i^2}{2}\le[H''(x_i)-\Delta H(x_i)\ri]+\frac{x_i^3y}{6}H'''(\xi_y x_i),
$$
for some
$\xi_y\in[1\wedge(1+y),1\vee(1+y)]$,
implies that
\begin{eqnarray*}
K_3(x_i) &\leq& \frac{x_i^3}{3!}\cdot \|H'''\|_{\infty} 
\cdot \int_{A_i^0(h)}|y|^3 g(x_i,y)\td y \leq 
 \frac{u^2}{3!}\cdot C'\cdot \|H'''\|_{\infty}\cdot h(n)(C_+(h(n)) + C_-(h(n)) + 2).
\end{eqnarray*}

Putting everything together we find that
\begin{eqnarray}
\label{eq:en}
\e'_{n}(H) &\leq& h(n)A_0\le[\|H'''\|_\infty 
+ \|H''\|_\infty+  \|H'\|_\infty + \|H\|_\infty\ri]  + 2k(n)\|H\|_{\infty}\\
&+& h(n)
\begin{cases}
A_1,  & \text{cases 0 and II},\\
A_2 +A_3|\log(h(n))|, & \text{case I,}
\end{cases}
\nn
\end{eqnarray}
%
%
%
for some positive constants
$A_i$,
$i=0,\ldots,3$.
All supremum norms  in \eqref{eq:en} are finite, 
since they are equal to the maxima of 
continuous functions over compact sets.
Combining \eqref{eq:en} with
Lemma~\ref{lem:gen-estimate0}, which is applicable in view of the assumptions,
the theorem follows. 
\exit